%% file: main.tex
\documentclass{article}
\usepackage{arxiv}

\usepackage[utf8]{inputenc} 
\usepackage[T1]{fontenc}    
\usepackage{hyperref}       
\usepackage{url}            
\usepackage{booktabs}       
\usepackage{amsfonts}       
\usepackage{nicefrac}       
\usepackage{microtype}      
\usepackage{makecell}
\usepackage{siunitx}
\usepackage{caption}
\usepackage{subcaption}
\usepackage{multirow}
\usepackage{xr}
\usepackage{cleveref}
\usepackage{graphicx}

\usepackage{comment}

\usepackage{xcolor}
\usepackage{subcaption}

\usepackage{multirow}
\usepackage{tablefootnote}

\usepackage{lipsum}
\usepackage{etoolbox}
\usepackage{setspace} 
\usepackage{tabularx}

\definecolor{blue}{RGB}{0,0,0}

\newcommand*\rot[1]{\hbox to4em{\rotatebox[origin=bl]{45}{#1}}}

\AtBeginEnvironment{quote}{\par\singlespacing\small}

\AtBeginDocument{%
  \providecommand\BibTeX{{%
    \normalfont B\kern-0.5em{\scshape i\kern-0.25em b}\kern-0.8em\TeX}}}

\title{Discovering the Hidden Facts of User-Dispatcher Interactions via Text-based Reporting Systems for Community Safety}

\author{
  Yiren Liu \\
  Informatics \\
  University of Illinois at Urbana-Champaign\\
  \texttt{yirenl2@illinois.edu} \\
   \And
  Ryan Mayfield \\
  LiveSafe, Inc.\\
  \texttt{ryan@livesafemobile.com} \\
  \And
    Yun Huang \\
  School of Information Sciences \\
  University of Illinois at Urbana-Champaign\\
  \texttt{yunhuang@illinois.edu} \\
}

\begin{document}

\maketitle


\begin{abstract}

Recently, an increasing number of safety organizations in the U.S. have incorporated text-based risk reporting systems to respond to safety incident reports from their community members. 
\textcolor{blue}{To gain a better understanding of the interaction between community members and dispatchers using text-based risk reporting systems, this study conducts a system log analysis of \textit{LiveSafe}, a community safety reporting system, to provide empirical evidence of the conversational patterns between users and dispatchers using both quantitative and qualitative methods. 
We created an ontology to capture information (e.g., location, attacker, target, weapon, start-time, and end-time, etc.) that dispatchers often collected from users regarding their incident tips. 
Applying the proposed ontology, we found that dispatchers often asked users for different information across varied event types (e.g., \textit{Attacker} for \textit{Abuse} and \textit{Attack} events, \textit{Target} for \textit{Harassment} events).
Additionally, using emotion detection and regression analysis, we found an inconsistency in dispatchers' emotional support and responsiveness to users' messages between different organizations and between incident categories.}
The results also showed that users had a higher response rate and responded quicker when dispatchers provided emotional support. 
These novel findings brought significant insights to both practitioners and system designers, e.g., AI-based solutions to augment human agents' skills for improved service quality.

\end{abstract}



\maketitle
 
\input{1_intro}

\input{2_related_works}

\input{3_method}
\input{4_results_RQ1}

\input{4_results_RQ2}

\input{4_results_RQ3}

\input{5_discussion}

\input{6_limitation}

\input{7_conclusion}


\bibliographystyle{ACM-Reference-Format}
\bibliography{sources}

\input{appendix}

\end{document}
\endinput

%% file: 1_intro.tex
\section{Introduction}

With the recent advance in community ICTs, an increasing number of organizations is adopting new technologies to help improve the effectiveness and efficiency of risk management. For example, text-to-911 allows users to communicate with dispatchers by sending text messages in addition to making phone calls \cite{goldstein2018next, grace2021text}. 
Many recent HCI works focus on improving the communication from community members to safety agencies by allowing users to report risk concerns in a crowd-sourcing manner \cite{blom2010fear, ahmed2014protibadi, cvijikj2015towards}. Increasingly, organizations have added text-based live chat safety incident reporting functionalities to their risk management systems \cite{uicSAFE, ming2021examining, wetip}. Studies have shown that incorporating live chat features in community risk management can not only increase users' perceived utility and reduce cost \cite{yabe2017cost}, but also benefit marginalized groups in risk reporting such as deaf and hard-of-hearing (DHH) and disabled people \cite{Textto911WhatYouNeedToKnow, ellcessor2019call}.

Despite the fact that plenty of studies have discussed the advantage of such systems, no studies have empirically investigated community members' and dispatchers' interaction behavior and emotional dialogue using text-based reporting systems. 
Recent research conducted a log analysis of university safety organizations' use of text-based system reporting through a mobile safety reporting app \cite{ming2021examining}, which provided insights over reporters' interactions with the system when submitting incident tips using the app. 
The findings of this study focused on organizations' different configurations of the app features and varied distributions of incident types without diving deep into the conversations, nor about how staff from organizations provide emotional support to their members. However, prior studies have strengthened the importance for dispatchers to offer emotional support in order to provide effective and efficient services \cite{clawson2018litigation, tracy1998emotion, feldman2021calming}. \textcolor{blue}{Further analysis needs to be done in order to study and comprehend detailed conversation interactions between dispatchers and users of such text-based reporting platforms.}

\textcolor{blue}{In this study, we investigate the conversational log of the LiveSafe risk reporting system.} We focus on analyzing the interactions between two stakeholders, community member users and dispatchers, during incident reporting live-chat conversations. We refer to the community members who submit reports as "Users" and human agents from safety organizations who handle reports during the chat sessions as "Dispatchers" throughout this paper. 
We examined in total 3,124 events that happened between 2018-2019 across 111 different safety departments of colleges and universities. 
\textcolor{blue}{We conducted quantitative and qualitative analysis to understand: what information was included when users sent tips, how dispatchers serviced the users' reports, and how users had follow-ups with the dispatchers.}

\textcolor{blue}{This study makes significant contributions for future HCI research in the area of text-based reporting systems for community safety.}
\textbf{First}, we conducted a conversational analysis based on empirical data collected from the application of a text-based risk reporting system in a real-life scenario, which yielded insights into the conversational interactions between users and dispatchers that are not discussed by prior studies. Specifically, we reviewed and annotated a sample of event chat logs with our proposed event information ontology.
By conducting further analysis, we found that users' reporting behavior exhibited significant differences according to tip categories.
We also discovered that users reporting anonymously tend to disclose more event-related information during reporting. 
\textbf{Second}, we identified an inconsistency in dispatchers' behavior of providing emotional support to users. A logistic regression analysis was conducted to reveal that the likelihood of a tip receiving emotional support from the agent was positively associated with certain event categories (i.e., \textit{MentalHealth}, \textit{EmergencyMessage} and \textit{TheftLostItem}). \textcolor{blue}{We also found that there were less emotional support delivered for those reports that were made on behalf of someone else or anonymously.}
\textbf{Third}, we found that users' responsiveness during the conversation with dispatchers was positively associated with factors related to service quality (i.e., if emotional support was provided, and dispatchers' response time). 
\textbf{Lastly}, our findings offer empirical evidence for understanding how the quality of services offered by safety agencies can influence users' behavior during the incident reporting process.
We draw implications from these findings to help improve designs of future text-based risk reporting systems and suggest novel designs for utilizing  conversational agents to assist safety agencies with providing effective and efficient community risk management services. 








%% file: 2_related_works.tex
\section{Related Work}
In this section, we present prior works related to community risk reporting systems. While we consider risk reporting beyond just crime, the majority of existing scholarship focuses on individual interactions with emergency systems like 911.
First, we discuss existing systems and studies designing text-based risk reporting systems. 
Second, we provide an overview of studies with a focus on the behavior of risk reporting system users and administrators respectively. We approached users' reporting behavior from the perspective of a utility-cost model, then we discussed existing works on how emergency line call-takers handle incident tips.

\subsection{Community Risk Reporting Systems}

Recent work has advanced the usage of Information and Communication Technology (ICT) for risk management in public domains \cite{sivvcevic2020possibilities}. 
For community risk reporting, many existing systems have been designed with different features in order to assist users in reporting safety events and incidents easily. 
In earlier efforts, web-form \cite{iriberri2006reporting} or mobile device \cite{sakpere2015usable} based reporting systems were introduced, providing alternative methods for users to report crimes and safety incidents.
Additionally, systems that enable users to tag and report safety concerns and risks geographically via crowd-sourcing have been designed and studied \cite{blom2010fear, cvijikj2015towards}. Later studies managed to connect emergency service providers (e.g., 911, medical service) with location-based incident reporting systems \cite{hossain2018bangladesh, kalyanchakravarthy2014android, ahmed2014protibadi}. These systems increase the effectiveness of users getting assistance by giving users quick access to emergency services and automatically sharing users' locations with emergency service providers. 



Many organizations have recently incorporated live-chat safety incident reporting features as part of their risk management systems \cite{uicSAFE, ming2021examining, wetip}. Prior study \cite{yabe2017cost} pointed out that campus community safety apps with emergency texting features offered a higher perceived utility by community members and a lower cost compared to traditional emergency communication systems such as blue-light emergency phone towers\cite{reaves2015campus}. Additionally, text-based reporting systems like text-to-911 were found to be beneficial to deaf and hard-of-hearing (DHH) and disabled people \cite{Textto911WhatYouNeedToKnow}. 
Albeit existing discussion over text-based reporting systems, current research lacks an empirical understanding of how users interact with safety administrative agents when using the aforementioned text-based reporting systems, especially during conversation. 


\subsection{Utility of Reporting --- Intrinsic and Extrinsic Factors}

Many existing works have discussed the issue of under-reporting, particularly in the context of reporting crimes using emergency phone lines (911). Myers \cite{myers1980crimes} examined under-reporting from the perspective of behavioral economics, suggesting the process of individuals determining whether to file police reports can be described with a utility-cost decision-making model \cite{bowles2009crime}. The decision of whether or not to report a safety incident or crime can be influenced by multiple factors. Inspired by theories of motivation \cite{wertheimer2020brief} and externality \cite{pigou2017economics}, we divide factors influencing crime reporting into two categories: intrinsic and extrinsic factors.

Some of them are intrinsic psychological factors of the reporter. Studies have shown that users' perceived confidence in police efficacy has a negative effect on their decision to report varied incidents including property crimes \cite{goudriaan2004reporting} and violence \cite{baumer2002neighborhood, anderson2000code}. Reporters of crimes were also found to have received intrinsic benefits resulting from  altruistic motives such as solidarity and incapacitation of offenders.

Extrinsic factors influencing reporting decisions are also multi-faceted. 
Community members' risk-reporting behavior is influenced by social factors. Social control theory \cite{hirschi2017causes} has been utilized by prior studies to understand the relationship between social interactions, risky acts, and informal enforcement of social norms. Studies suggest social components are a factor influencing individual crime reporting decisions. Goudriaan et al. \cite{goudriaan2006neighbourhood} approached crime reporting from the perspective of social cohesion, suggesting that as a crucial factor for realizing informal social control  \cite{silver2004sources}, social cohesion has a positive effect on the likelihood of community crime reporting. 
Another factor suggested by previous works is socio-economic condition. Works have shown that people living in socio-economically disadvantaged neighborhoods are less likely to report crimes \cite{goudriaan2006neighbourhood, myers1980crimes, baumer2002neighborhood}. The study also showed that people with higher income and education status are more likely to report violence and property crimes \cite{xie2014area}. 
Other external factors include the amount of time needed to describe the incident to police and maybe testify in court, as well as potential offenders' retaliation \cite{ziegenhagen1977victims}. 

Although plenty of studies have discussed the potential factors influencing people's decision-making in crime reporting, no empirical evidence has been provided to demonstrate the effect of potential factors on users' behavior during reporting. 
Gray et al. \cite{gray2002dealing} discussed varied factors (e.g., confidence in the agency and perceived uncertainty) and their impact on users' perception of risk. According to the study, individuals experience an increased perception of fear when they have less trust in the organization handling the risky event. This affects individual behavior while reacting to the incident, including engagement during reporting, even after they have decided to report the incident.
Based on the utility-cost model of reporting, we want to investigate the factors that associate with people's reporting behavior after they decide to report an incident, and provide implications to improve the efficiency and effectiveness of the reporting process.



\subsection{Empirical Understanding of Community Engagement During Risk Reporting}

Previous works sought to understand and improve the interaction between emergency dispatchers and callers, both from an emotional perspective and an organizational perspective. 
In practice, users reporting an incident might be subject to different causes that could lead to emotional instability. 
A previous observational study showed that users who are experiencing or have experienced difficult or traumatizing safety incidents will often exhibit "emotional pain" \cite{Observations_on_the_Display}. The work stated that it is the dispatchers' responsibility to handle the emotional display of users to increase the users' trust in dispatchers and willingness to cooperate. Meanwhile, several strategies used by dispatchers were observed including using directives, reassurances, and realignment.
Paoletti \cite{paoletti2012operators} also discovered that the questioning sequence asked by emergency call takers could also be perceived by users as 'inappropriate and a way of delaying assistance', thus leading to annoyance and anger. 
Feldman \cite{feldman2021calming} analyzed the methods used by dispatchers to handle callers' emotional outbursts using conversation analysis based on the emergency call logs. In this work, strategies beyond those discovered in Whalen et al.'s work \cite{Observations_on_the_Display} were identified including repetitive persistence and redirection of attention.
Although these works discussed the strategies used for responding to callers' emotions, no work has yet investigated such strategies adopted in the context of text-based risk reporting systems.

Meanwhile, for reports, also called tips, made by users using mobile or web-based applications, the issues mentioned in these reports were often found to be less emergent than those reported through 911 emergency lines \cite{iriberri2006reporting}. Although the emotional aspect of incident reporting has been well discussed by prior studies, the previous accounts are limited to the context of emergency call-taking but not text-based reporting systems. 
Meanwhile, none of the prior work used real conversation records to identify the patterns of safety reporting under the context of chat-based safety tip reporting systems. 

In this study, we examined the complete conversational process between the user and the dispatcher, i.e., 1) The user submits an incident "tip" as the first message; 2) Based on the user's first message, the dispatcher asks questions to collect information and provide support; 3) The user provides additional information in response to the dispatcher's questions.
Based on the literature above, we would like to propose the research questions of this study as follows:
\begin{itemize}

    \item \textbf{RQ1}: What information is included (or missed) in users' first messages?


    \item \textbf{RQ2}: How do dispatchers interact with users to collect information and provide support? 
    


    \item \textbf{RQ3}: How do users engage in dispatchers' follow-ups?
    %
    

\end{itemize}

%% file: 3_method.tex
\section{Method}

\subsection{Dataset}
We utilize a dataset of real-world user-dispatcher dialogues from the domain of safety incident reporting. The dialogues were collected through the LiveSafe app across the time period between 2018 and 2019, which allows users to report certain events to a human agent via a chat service.

\begin{figure}[h]
  \centering
     \begin{subfigure}[b]{0.22\textwidth}
         \centering
         \includegraphics[width=\textwidth]{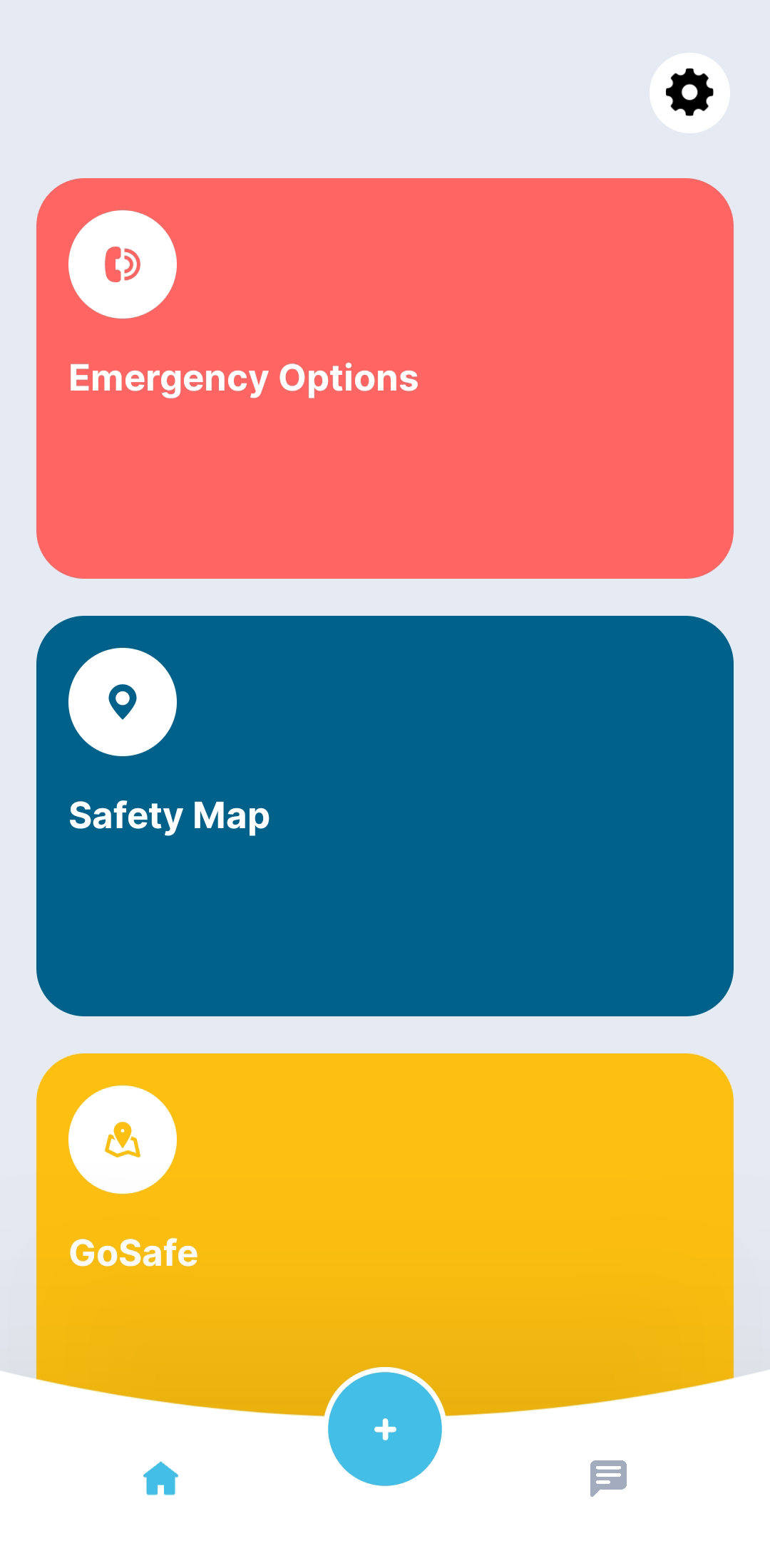}
         \caption{Home Menu}
         \label{fig:livesafeInterface: HomeMenu}
     \end{subfigure}
     \begin{subfigure}[b]{0.22\textwidth}
         \centering
         \includegraphics[width=\textwidth]{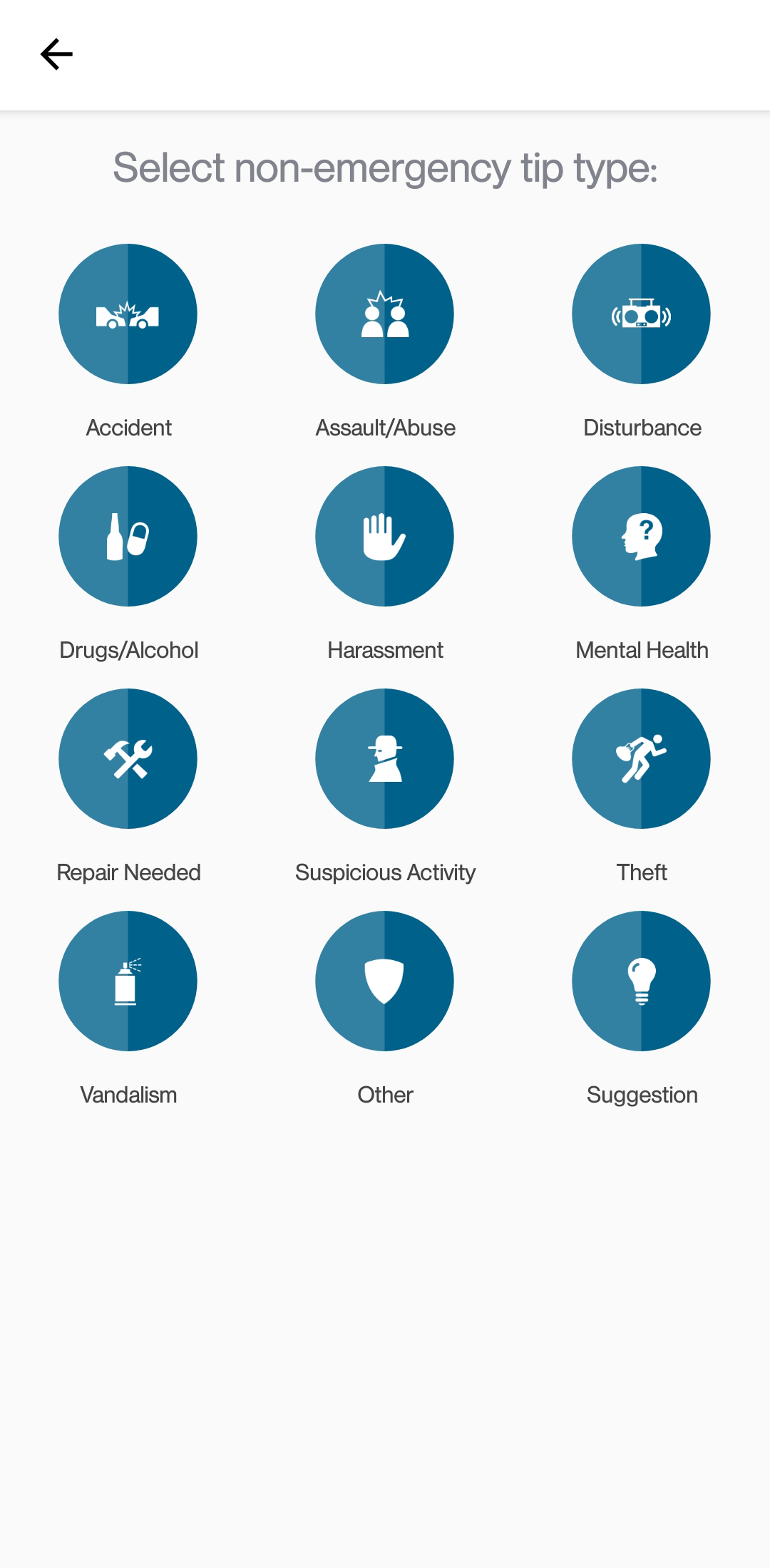}
         \caption{Tip Categories}
         \label{fig:livesafeInterface: tipCategory}
     \end{subfigure}
     \begin{subfigure}[b]{0.22\textwidth}
         \centering
         \includegraphics[width=\textwidth]{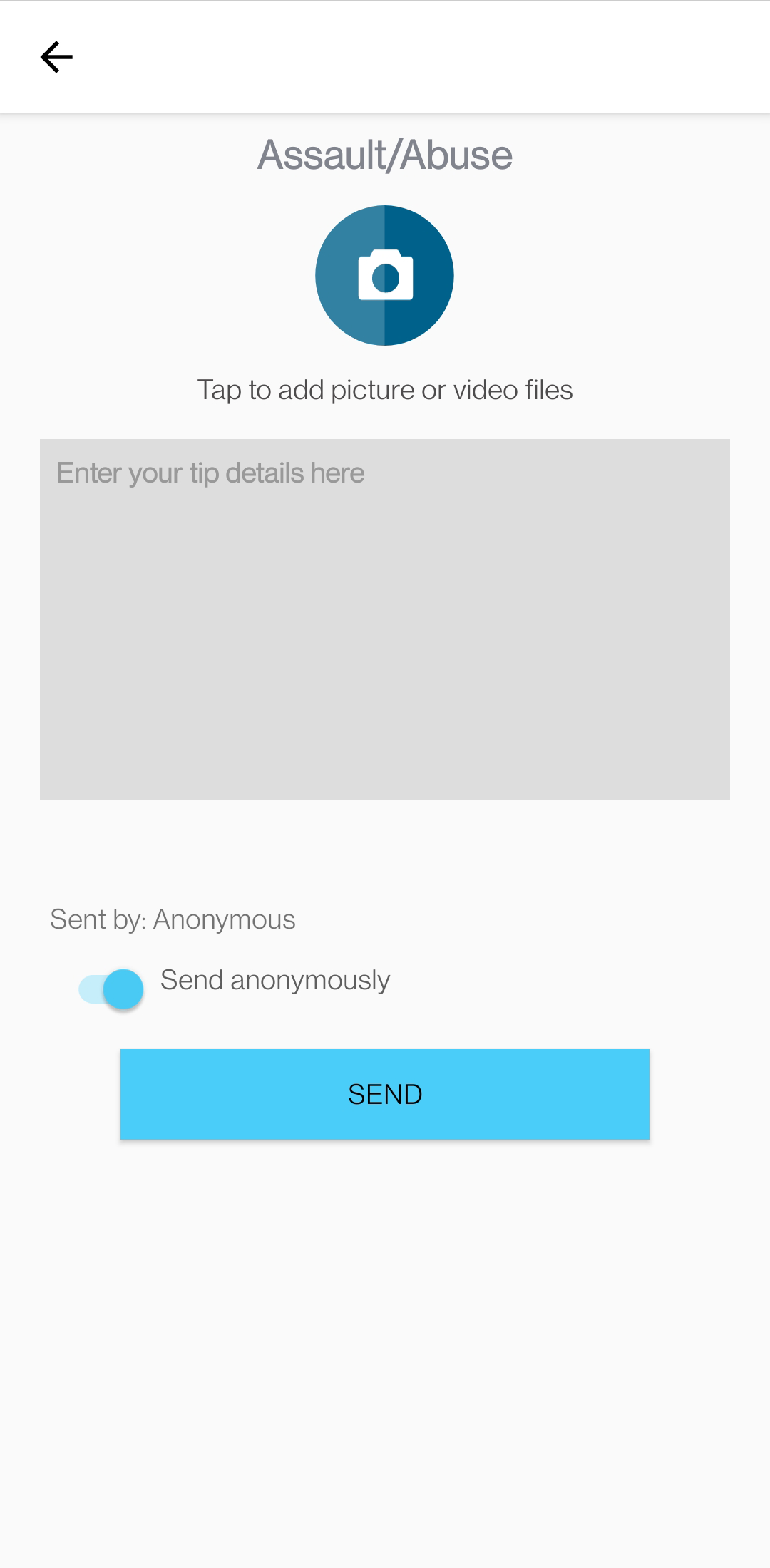}
         \caption{Submitting Tip}
         \label{fig:livesafeInterface: submitTip}
     \end{subfigure}
     \begin{subfigure}[b]{0.22\textwidth}
         \centering
         \includegraphics[width=\textwidth]{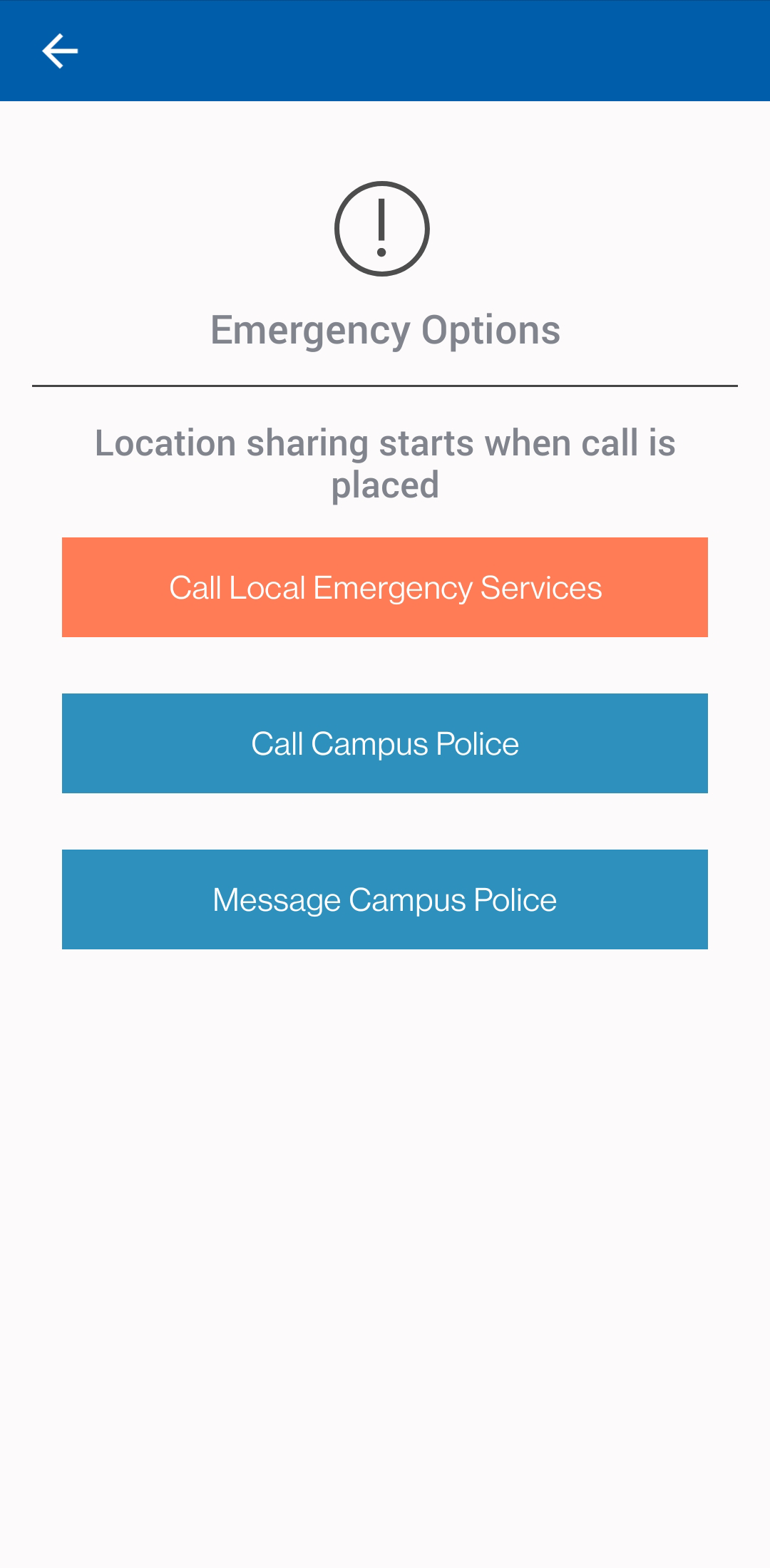}
         \caption{Emergency Options}
         \label{fig:livesafeInterface: emergencyOptions}
     \end{subfigure}
     \\
     \vspace{0.5cm}
      \begin{subfigure}[b]{0.85\textwidth}
         \centering
         \includegraphics[width=\textwidth]{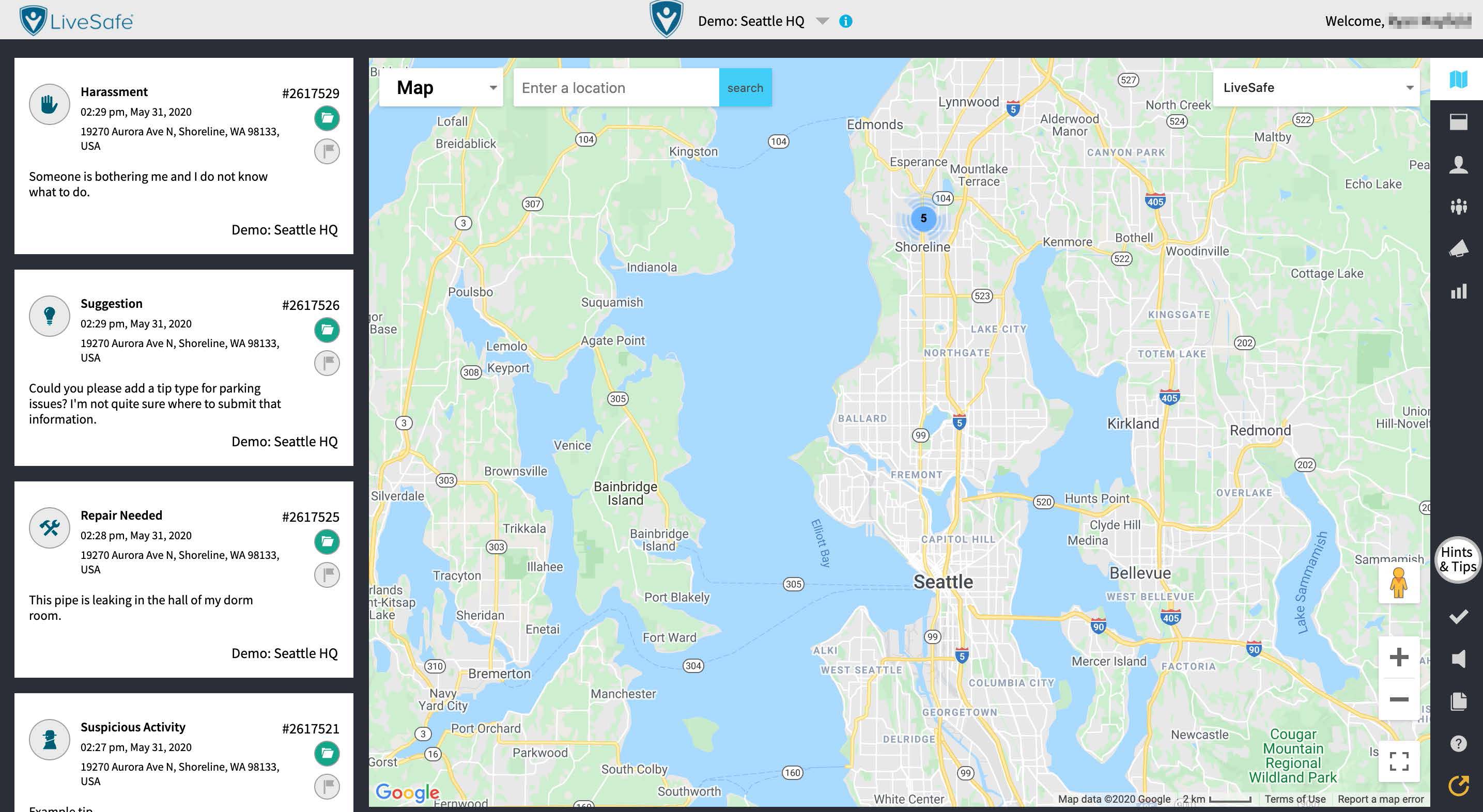}
         \caption{Dispatcher View}
         \label{fig:livesafeInterface: dispatcherView}
     \end{subfigure}
  \caption{The user interface of the LiveSafe mobile app; a) The user can choose to submit a tip or enter emergency options; b) The user needs to select an incident category after choosing to submit a tip; c) After the incident category is chosen, the user can submit text and image description and connect to a human agent; d) The user can also choose to dial emergency number from the home menu in case of emergency; e) The dispatcher can connect and chat with the user using the Command Dashboard.}
  \label{fig:livesafeInterface}
\end{figure}

LiveSafe is a community-based risk management system that focuses on increasing involvement and communication between members of an organization and the safety teams responsible for mitigating broad-spectrum risk \cite{ming2021examining}. This tool was first introduced in 2013, and has since been used by over 200 higher education institutions. Community members submit information ("tips") through the LiveSafe mobile app or web portal, while safety organization managers respond to tips through the LiveSafe Command Dashboard. 
Users can choose from a variety of "tip types" and include text, photographs, video, or audio recordings in their tips when submitting them. The local safety departments choose the setup of tip types for each organization, as shown in Fig. \ref{fig:livesafeInterface}. 
After the user's submission, the dispatchers from safety departments can reply to each tip and start a conversation with the user through the Command Dashboard provided by LiveSafe, as shown in Fig. \ref{fig:livesafeInterface: dispatcherView}. 

The initial dataset contains a large amount of noise, particularly in the form of niche or organizational-specific tip categories. 
We first cleaned up the data by removing "Test" tips and automatically created "Calls" entries (where users call the admins via the app), then we removed tips that were submitted within the first year of implementation for each organization to ensure the organization has formed a consistent report handling workflow. To further simplify the analysis, we only kept tips after 2018 and chats with more than 4 utterances (two conversation turns), since we want to investigate chats with at least one turn of information-collecting conversation. 
We then calculated the average chat utterance count for each event category and selected tips under the top-6 event categories.
Finally, for each tip we performed language detection using \textit{langdetect} \cite{nakatani2010langdetect} to identify messages in English. We identified 17.0\% (690 out of 4053) of the tips to be non-English, and removed them to simplify further analysis. The considerable amount of non-English tips also suggests a potential need for multilingual support for text-based risk reporting systems, particularly in those organizations that carried a high amount of non-English submissions. 



As shown in Table \ref{Tab: dataset stats}, after performing cleaning, the resulting dialogues are categorized into different incident categories, which individuals selected before initiating a chat conversation. Sensitive information (names, places, times, etc) in the dataset was masked for privacy. 
It is worth noting the impacts this might have had on the annotation process and the model performance, as it is not always clear what the masked value was (e.g. "The incident must have occurred sometime between [TIME] and [TIME] which is when I got out of work.").

\begin{table}[!ht]
\resizebox{\linewidth}{!}{
\begin{tabular}{cccccc}
\hline
\textbf{\begin{tabular}[c]{@{}c@{}}Tip \\ Category\end{tabular}} & \textbf{\begin{tabular}[c]{@{}c@{}}Dialogue \\ Number\end{tabular}} & \textbf{\begin{tabular}[c]{@{}c@{}}Utterance \\ Number\end{tabular}} & \textbf{\begin{tabular}[c]{@{}c@{}}Avg. \\ Word \# \\ per Dialogue\end{tabular}} & \textbf{\begin{tabular}[c]{@{}c@{}}Avg. \\ Word \# \\ per Utterance\end{tabular}} & \textbf{\begin{tabular}[c]{@{}c@{}}Avg. \\ Utterance \# \\ per Dialogue\end{tabular}} \\ \hline
SuspiciousActivity                                               & 1164                                                                & 10347                                                                & 92.15                                                                              & 10.37                                                                               & 8.89                                                                                      \\
EmergencyMessage                                                 & 807                                                                 & 7190                                                                 & 77.19                                                                              & 8.66                                                                                & 8.91                                                                                      \\
DrugsAlcohol                                                     & 689                                                                 & 5447                                                                 & 70.62                                                                              & 8.93                                                                                & 7.91                                                                                      \\
HarassmentAbuse                                                  & 299                                                                 & 2817                                                                 & 103.70                                                                             & 11.01                                                                               & 9.42                                                                                      \\
TheftLostItem                                                    & 280                                                                 & 2538                                                                 & 84.50                                                                              & 9.32                                                                                & 9.06                                                                                      \\
MentalHealth                                                     & 124                                                                 & 1355                                                                 & 112.20                                                                             & 10.27                                                                               & 10.93                                                                                     \\ \hline
\end{tabular}
}
\caption{Descriptive statistics by tip category}
\label{Tab: dataset stats}
\end{table}


The dataset provides abundant conversation logs for further analysis. To distinguish it from the later annotated sample, we refer to this dataset as all data, containing a total of 3363 conversations and 29,694 utterances.



\subsection{Event Ontology and Annotation}
To capture the event information from the conversations between users and dispatchers, we established an event ontology for community event extraction. We refer to ontology as the same concept in dialog state tracking task \cite{budzianowski2018multiwoz}. The ontology is based on the event ontology of Automatic Content Extraction (ACE) \cite{walker2006ace} dataset.  

To simplify the annotation process, we focus on two of these categories - \textit{TheftLostItem} and \textit{HarassmentAbuse}. These categories have the longest number of exchanged utterances. We chose the top 3 tip categories set up by the highest percentage of organizations as user-selectable options when submitted tips, i.e. \textit{HarassmentAbuse} (98.6\%), \textit{SuspiciousActivity} (97.2\%), and \textit{TheftLostItem} (92.9\%). Since Suspicious Activity category is comprised of a mixture of events with obscured event types \cite{ming2021examining}, we selected \textit{HarassmentAbuse} and \textit{TheftLostItem} as two main categories of our analysis.

Two researchers independently sampled 10 dialogues from each category and proposed an ontology draft, then merged the two drafts. The two researchers then annotated a small sample of dialogues and discussed frequently until the annotation converged. 

The final proposed ontology consists of three parts: event type, event argument and dispatcher intent. Since the initial incident category information provided by the LiveSafe dataset is coarse-grained and incident tips within the same category could turn out to be completely different events in natural (e.g. domestic abuse and random harassment under HarrassmentAbuse) \cite{ming2021examining}, we manually annotated more fine-grained event type information in ordering to distinguish such events. The event types from our ontology include \textit{Attack}, \textit{Harassment}, \textit{Abuse}, \textit{Threat}, \textit{Robbery}, \textit{Theft} and \textit{Break-in}. Event types \textit{Attack}, \textit{Harassment}, \textit{Abuse} and \textit{Threat} belong specifically to category \textit{HarassmentAbuse}, and \textit{Robbery}, \textit{Theft} and \textit{Break-in} to \textit{TheftLostItem}. During annotation of each event, we selected a word or phase as the event handle and annotate the event type over it. 
For the annotation of event arguments, we refined the event argument schema based on the Automatic Content Extraction (ACE) dataset \cite{walker2006ace} and adapted it to the domain of safety incident reporting. The ACE dataset is an event extraction task dataset widely used by prior studies \cite{lin2020joint, xia2019multi, li2021document}. The dataset contains annotations about event and entity information from news articles. The dataset partially covers events adaptable to the safety domain such as \textit{Attack} with arguments including \textit{Attacker}, \textit{Target}, \textit{Instrument}, \textit{Time}, and \textit{Place}. The original ACE dataset schema, though, is too coarse-grained to be used for directly annotating safety-related incidents. As a result, we modified the ACE event schema and applied it to analyze LiveSafe chat logs. 

\begin{figure}[h!]
  \centering
  \includegraphics[width=.9\linewidth]{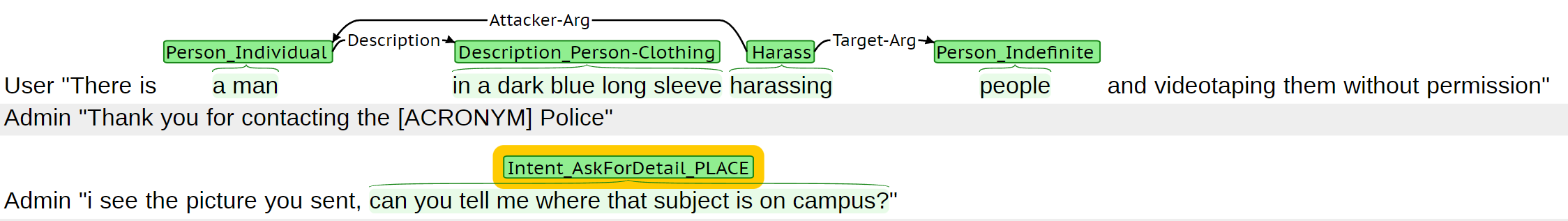}
  \caption{Example Snippet from an Annotated Dialogue}
  \label{fig:annotation_example}
\end{figure}

The event arguments we annotated include \textit{Attacker}, \textit{Target}, \textit{Location}, \textit{Weapon}, \textit{Start Time}, \textit{End Time}, and \textit{Target Object}. Each event argument is connected to the event handle, as a part of the event information. 
We also annotated dispatchers' intent when they asked users questions to collect additional event-related information. We noted which event argument dispatchers were asking for more information about. For example, as in Fig. \ref{fig:annotation_example}, the dispatcher asked "where that subject is on campus", which was annotated as asking for detail about \textit{Location} information. 
The complete ontology and definition can be found in Appendix \ref{appendix}.

After the final ontology was proposed, two researchers annotated a total of 69 event chat logs with the new ontology. The result annotations are considered consistent given an inter-annotator agreement (Cohen’s Kappa) of 0.87. The annotation process is completed using brat \cite{brat}. An example of annotated dialogue is shown in Fig. \ref{fig:annotation_example}. We refer to the resulting annotated dataset as the annotated sample data, containing a total of 69 conversations and 972 utterances.


\subsection{Detecting Emotional Support and Reporting on Behalf or as Witness}
\label{sec: method_t5}
We used NLP techniques to identify two attributes for each conversation: whether or not the dispatcher provided emotional support (defined as "expressing concern, affection or caring, providing comfort or encouragement" \cite{vaux1987modes}), and whether or not the user was reporting on behalf of or as a witness. This allowed us to better understand the variables that could affect users' and dispatchers' responding behavior.

First, to discover emotional displays from dispatchers throughout the chat, we implemented an emotion classifier using the emotion classification dataset from Saravia et al.'s work \cite{saravia2018carer}. 
Previous work used an LSTM-based model to identify emotional support within the domain of Online Health Communities \cite{khanpour2018identifying}.
To further improve the context-awareness and domain adaptability of the classification model, we finetuned the T5 language model with the emotion dataset using the text classification task setting as described in \cite{raffel2019exploring}. The resulting model was able to classify text into its corresponding emotion categories including \textit{anger}, \textit{fear}, \textit{joy}, \textit{sadness}, \textit{surprise} and \textit{love}. The model is evaluated on the test set provided by the same work, reaching a satisfying overall accuracy of 93\% on the GoEmotion Dataset.
We then apply the classification model overall chat logs within the dataset. In order to identify the cases where the dispatcher provided emotional support to the user, we picked out events where at least one of the dispatcher's responses was classified into the category of \textit{sadness} (the model achieved an evaluation accuracy of 96\% on the GoEmotion Dataset for category \textit{sadness}), which resulted in a total of 191 events.
\textcolor{blue}{
We followed the methods used in prior studies \cite{chakraborty2017tabloids, bagroy2017social}, to further validate the model's accuracy over our own dataset. Specifically, we randomly sampled 100 utterances from the predicted chats. One researcher then annotated the emotions of these utterances as ground truth, and then compared them to the model predicted labels for evaluation. The results yielded  82\% accuracy for the sampled 100 utterances, including all predicted emotion types. 
We also sampled and annotated 100 utterances randomly from chats that are predicted as \textit{sadness}. The model achieved an accuracy of 86\% within the sample from chat utterances predicted as \textit{sadness}.
}


To determine whether or not a tip is reported by the victim, we identify tips where users explicitly mentioned that they were reporting on behalf or as a witness.
To achieve this, we search for keywords ``friend'' or ``victim'' within the chat log of each tip. This is due to the fact that during the annotation process, we discovered that the majority of users who were witnesses or reporting on behalf of friends would explicitly state that they were doing so, or the dispatcher would ask the user for details about the victim.
This information was then constructed as a categorical variable during the logistic regression. 

%% file: 4_results_RQ1.tex
\section{Findings}
In this section, we present the analysis results to address the proposed research questions. 
First, we discuss what information is provided in the first message by the user upon submitting the incident tip, and what information is initially missed and later elicited by the dispatcher. 
Second, we investigate the strategies used by dispatchers to collect incident-related information from users and provide emotional support.
Third, we study the potential factors influencing users' responsiveness during conversations with dispatchers.
A summary of the major findings of each RQ is also provided at the end of each subsection. 
\subsection{What Information is Often Reported (or Missed) by the Users? (RQ1)} 


By annotating a sample of data using the proposed event ontology, we aim to understand the event distribution across the sampled chat log dataset. 
The annotation covers information including fine-grained event types, event arguments, and dispatcher intents. Event arguments include essential details that are needed to be identified from the dialogue between the dispatcher and the user to capture the event information. 

The resulting dataset after annotation consists of in total 69 incidents, including 42 HarrassmentAbuse incidents and 27 TheftLostItem incidents. The fine-grained event types includes \textit{Harassment} (n=13), \textit{Attack} (n=10), \textit{Threat} (n=8) and \textit{Abuse} (n=11) under the HarrassmentAbuse domain; \textit{Theft} (n=17), \textit{Robbery} (n=6) and \textit{Break-In} (n=4) under the TheftLostItem domain. 

Using the proposed ontology, we also annotated the event arguments mentioned in the conversations. As shown in Table \ref{Tab: args_annotated} for both HarassmentAbuse and TheftLostItem, the event argument type with the highest number of mentions is \textit{Location}, which is crucial information needed to file an incident report.

\begin{table}[h!]
\resizebox{\linewidth}{!}{
\begin{tabular}{ccccccccc}
                                 &                & \multicolumn{7}{c}{\textbf{Event Argument}}                                                                                                     \\ \cline{3-9} 
                                 &                & \textbf{Location} & \textbf{Attacker} & \textbf{Target} & \textbf{End\_Time} & \textbf{Weapon} & \textbf{Target\_Object} & \textbf{Start\_Time} \\ \hline
\multirow{5}{*}{\rotatebox[origin=c]{90}{\parbox[c]{2cm}{\centering Harrassment-Abuse}}} & Abuse          & 19                & 6                 & 9               & 3                  & 1               &                         &                      \\
                                 & Attack         & 11                & 12                & 4               & 4                  & 1               &                         &                      \\
                                 & Harassment         & 7                 & 12                & 13              & 3                  &                 &                         &                      \\
                                 & Threat       & 5                 & 6                 & 7               &                    &                 &                         &                      \\
                                 & \textbf{Total} & \textbf{42}       & \textbf{36}       & \textbf{33}     & \textbf{10}        & \textbf{2}      & \textbf{}               & \textbf{}            \\ \hline
\multirow{4}{*}{\rotatebox[origin=c]{90}{\parbox[c]{1.5cm}{\centering Theft-LostItem}}}   & Break-In       & 4                 & 1                 &                 & 1                  &                 &                         &                      \\
                                 & Robbery         & 6                 & 2                 & 2               & 2                  &                 & 2                       &                      \\
                                 & Theft         & 22                & 9                 & 5               & 9                  &                 & 24                      & 2                    \\
                                 & \textbf{Total} & \textbf{32}       & \textbf{12}       & \textbf{7}      & \textbf{12}        & \textbf{}       & \textbf{26}             & \textbf{2}           \\ \hline
\end{tabular}
}
\caption{Distribution of Event Arguments by Event Type within Annotated Incidents}
\label{Tab: args_annotated}
\end{table}

\textbf{How Much Information is Provided in the First Place?}
During an incident reporting dialogue using the LiveSafe app, the user was required to first submit a tip as the first message of the conversation describing the details of the incident. 
We calculate the average percentage of event arguments mentioned at the beginning of conversations (incident tips) compared to that mentioned during the complete conversations, to measure how much information is provided by the user without dispatchers involved in the first place. 

\begin{table}[h!]
\resizebox{\linewidth}{!}{
\begin{tabular}{cccccccc}
\hline
\textbf{Incident Category}                                                                    & \multicolumn{4}{c}{\textbf{HarassmentAbuse}}                             & \multicolumn{3}{c}{\textbf{TheftLostItem}}            \\ \hline
\textbf{Event Type}                                                                           & \textbf{Abuse} & \textbf{Attack} & \textbf{Harassment} & \textbf{Threat} & \textbf{Break-In} & \textbf{Robbery} & \textbf{Theft} \\ \hline
\begin{tabular}[c]{@{}c@{}}avg. \\ entity count \\ at the beginning
\end{tabular} & 2.73 (1.13)    & 2.90 (1.7)      & 4.69 (1.72)         & 2.50 (1.32)     & 1.50 (0.87)       & 2.50 (0.96)      & 3.88 (1.45)    \\ \hline
\begin{tabular}[c]{@{}c@{}}avg. \\ entity count \\ after dispatchers engaged\end{tabular}   & 4.64 (1.55)    & 6.30 (2.10)     & 6.31 (2.61)         & 4.12 (1.96)     & 1.75 (1.30)       & 3.67 (2.49)      & 5.53 (2.35)    \\ \hline
avg. percentage increased                                                                     & 97\% (103\%)   & 161\% (236\%)   & 34\% (38\%)         & 129\% (224\%)   & 8\% (14\%)        & 43\% (51\%)      & 50\% (62\%)    \\ \hline
\end{tabular}
}
\caption{Average count of entities revealed at the beginning/end of conversations and percentage of increase; Numbers in parentheses
are SD of entity counts}
\label{tab:entityRatioIncreased}
\end{table}


As shown in Table. \ref{tab:entityRatioIncreased}, for all event types, the average count of entities mentioned by users increased significantly after dispatchers are engaged in the conversations. 
For users' initial reports without dispatchers' intervention, there is no significant difference in the types of information that are mentioned. Information including \textit{attacker}, \textit{target}, \textit{location}, and \textit{time} is most often reported initially by users among all event types, except for events with type \textit{Theft},  \textit{object stolen} is most often mentioned. 
After agents finished collecting information from users, the types of information elicited differ by event type. For \textit{harassment} events, dispatchers are observed to have asked for targets' names and contact information (n=3). 

For \textit{Attack} and \textit{Abuse} events, we found that dispatchers would often further collect information about \textit{Attacker} from users, while information about \textit{Target} is most often collected from users during \textit{Harassment} events, as shown in Table \ref{arg_count_increased}.
For \textit{TheftLostItems} events, information about the \textit{object stolen} was also provided by users in the first place, and dispatchers most often asked for information about the \textit{attacker} (suspect).

\begin{table}[h!]
\centering
\begin{tabular}{cccccllc}
\hline
\textbf{} & \multicolumn{4}{c}{\textbf{HarassmentAbuse}}                                              & \multicolumn{3}{c}{\textbf{TheftLostItem}}                                                    \\ \hline
\textbf{} & \textbf{Abuse}       & \textbf{Attack}      & \textbf{Harassment}      & \textbf{Threat}    & \multicolumn{1}{c}{\textbf{Break-In}} & \multicolumn{1}{c}{\textbf{Robbery}} & \textbf{Theft} \\ \hline
Attacker  & 6 (8)                    & 12 (14)                   & 12 (4)                    & 6 (1)                    &     1                                  &   2 (2)            & 9 (5)               \\
Target    & 9 (4)                    & 4 (1)                    & 13 (7)                    & 7 (1)                    &                                       &  2 (1)           & 5 (2)               \\
Time      &  3                      & 4 (2)                    & 3 (2)                    &                        &          1 (1)                     &   2            & 11 (2)               \\
Location     &  19                     & 11 (2)                    &    7                 &  5 (4)                    &       4                                &     6                                & 22 (2)               \\
Target Object    &                   &                       &                      &                       &                                       &     2                                & 24 (1)               \\ 
Weapon    &      1                    &    1                   &                       &                       &                                       &                                     &                  \\ \hline
\end{tabular}
\caption{Event argument counts initially provided by users and elicited by dispatchers; \textbf{Numbers in parentheses are argument counts elicited by dispatchers during conversations}}
\label{arg_count_increased}
\end{table}


To further investigate users' incident reporting behavior, we conducted a one-way Analysis of Variance (ANOVA) test to examine the effect of event type on the ratio of event arguments mentioned in the user's first utterance. 
As shown in Table \ref{Tab:ANOVA_arg_ratio_firstMsg}, the ANOVA test revealed that when the event types are different, users reported event information statistically significant difference regarding the total count of event entities provided before dispatchers asked any questions ($F(6, 59) = [2.96], p = 0.01^{**}$). 
Further Tukey’s HSD Test revealed that the mean value of event entity number mentioned by users at the beginning was significantly different between \textit{Attack} and \textit{Threaten} ($p = 0.046^{**}, 95\% \ C.I. = [-5.99, -0.03]$), and \textit{Break-In} and \textit{Stolen} ($p = 0.04^{**}, 95\% \  C.I. = [0.12, 6.99]$). The results suggest that users involved in \textit{Break-In} events mentioned less information at the beginning of the conversation than those involved in \textit{Theft} and \textit{Attack} more than \textit{Threat} events.

To further reveal how much additional information was collected after dispatchers were involved, ANOVA results showed that within the incident category of \textit{HarassmentAbuse}, for different event types the total counts of additional event entities collected by dispatchers are statistically significantly different ($F(3, 36) = [3.26], p = 0.03^{**}$), as shown in Table \ref{Tab:ANOVA_entities_increased}.
Further Tukey’s HSD Test revealed that the mean value of event entities mentioned by users at the beginning was significantly different between \textit{Attack} and \textit{Threat} ($p = 0.04^{**}, 95\% \  C.I. = [-4.41, -0.08]$). The results suggest that dispatchers were able to elicit more event-related information from users when encountering \textit{Attack} events than \textit{Threat} events. 


\begin{table}[h!]
    \parbox{.48\linewidth}{
        \centering
\begin{tabular}{ccccc}
\hline
\textbf{}  & \textbf{SS} & \textbf{df} & \textbf{F} & \textbf{p-value} \\ \hline
Event Type & 97.76         & 6         & 2.96   & 0.01 (**)         \\ \hline
Residual   & 325.11         & 59        &            &                  \\ \hline
\end{tabular}
\caption{The ANOVA test of event type on the count of event entities mentioned in users' first utterance}
\label{Tab:ANOVA_arg_ratio_firstMsg}
    }
    \hfill
    \parbox{.48\linewidth}{
        \centering
\begin{tabular}{ccccc}
\hline
\textbf{}  & \textbf{SS} & \textbf{df} & \textbf{F} & \textbf{p-value} \\ \hline
Event Type & 26.77         & 3         & 3.25   & 0.03 (**)         \\ \hline
Residual   & 98.59         & 36        &            &                  \\ \hline
\end{tabular}
\caption{The ANOVA test of event type on the count of additional event entities collected by dispatchers}
\label{Tab:ANOVA_entities_increased}
        }
\end{table}

\textbf{What Information is Often Provided in Total for Each Type of Events?} 
To measure the comprehensiveness of events covered by each event argument, for each event argument we calculate the percentage of events mentioning that argument as the \textbf{coverage ratio}. The coverage ratio can be used to quantify how frequently each event argument is mentioned within a collection of incident reporting conversations (e.g., incidents of event type \textit{Abuse}). 

For \textit{Harassment} events, the \textit{Target} argument number has the highest coverage ratio (100\%) to event count, which indicates \textit{Target} information is almost always explicitly mentioned by users when reporting for \textit{Harassment} events. In contrast, users most often mention \textit{Location} argument when reporting for \textit{Abuse} (100\%), \textit{Break-In} (100\%) and \textit{Robbery} (83.3\%) events. 
Besides \textit{Harassment}, event type \textit{Threat} (87.5\%) and \textit{Abuse} (81.8\%) also have higher \textit{Target} argument coverage ratio comparing to \textit{Attack} (44.4\%). This might be the case because users were themselves targets of \textit{Attack} events, so the \textit{Target} information did not need to be explicitly mentioned. 

\textbf{Users Reporting on Behalf or as Witness} 
We also observed cases where the user is reporting on behalf of the actual victim, as in the following example:

\begin{quote}

\textbf{Incident category}: HarassmentAbuse;
\textbf{Event Type}: Harassment;
\textbf{Anonymous Tip}: Yes;

\vspace{0.2cm}

\textbf{User:} "my friend said an older guy kept grabbing her shoulder and trying to talk to her and wouldn't let her go on her own way. She was walking toward [LOCATION]"

\textbf{Dispatcher:} \textit{"Hello thank you for using LiveSafe, do you have a description. race clothing?"}

\textbf{User:} "she said possibly [RACE] in a black shirt with some facial scruff"

\textbf{User:} "some grey hairs which threw her off bc he said he was a current student"

\textbf{Dispatcher:} \textit{"ok, can you still see him or what direction he was last seen?"}

\textbf{User:} "she said she was on her way to [LOCATION] and he came outta nowhere by the bus station"

\textbf{Dispatcher:} \textit{"Ok we have notified officers and will check area, was your friend injured? Does she need medical attention?"}

\textbf{User:} "No! She is fine. She just said everytime she tried exiting the conversation he was grab her shoulder and continue the convo"

\textbf{Dispatcher:} \textit{"I'm very glad your friend is fine! We have officers checking the area for the male subject. Would she like to meet with officers?"}

\textbf{User:} "no i think she doesn’t want to. But thank u"

\end{quote}

The user, i.e., the victim's friend, is reporting the incident on behalf of the victim. In this case, the victim's information is explicitly stated by the user at the beginning of the conversation. The user also mentioned that the victim did not want to meet with officers when asked by the dispatcher. The victim might not want to report an incident directly. The friend or acquaintance could elect to report without the victim's permission or could have been asked to submit the report by the victim.
It is worth noting that, even though it was revealed that the user was reporting on behalf of the victim, the dispatcher persisted to ask if the user was the victim. The user reported anonymously but the dispatcher did not attempt to gather contact information from the user, which could make it difficult for the dispatcher to follow up with the user on the report.

By further examining other cases within our annotated sample, we found in total 24 out of 69 incidents (34.8\%) were reported by users that were not victims in their corresponding events. Among these cases, there existed 6 cases where users reported \textit{HarassmentAbuse} incidents (N=42) for victims for their friends. As for \textit{TheftLostItem} incidents (N=27), we also observed 2 cases where users reported theft for their friends. 
For the remaining incidents, we found 4 cases of domestic violence witnessed or overheard by users, and 12 cases where the report was made for incidents witnessed on the street.



\textbf{Does Anonymity Affect Users' Reporting Behavior?}
Out of all annotated incident reports, more than half of the reports (57.6\%) are submitted anonymously. More specifically, events with type \textit{Attack} (88.9\%), \textit{Robbery} (83.3\%), \textit{Harassment} (75.0\%) and \textit{Abuse} (72.7\%) had the highest percentages of anonymous reports, whereas \textit{Theft} (18.8\%), \textit{Break-In} (25.0\%) ranked the lowest. 

A further two-sample t-test revealed that the first messages (incident tips) reported by users anonymously have higher ratios of event information mentioned compared to non-anonymous reports ($F(28, 38) = [1.83], p = 0.04^{**}$). 


\hspace{0.01mm}

\textbf{Summary}: By analyzing the event-related information provided by users from the annotated sample, our findings revealed that users have different behavior patterns when reporting for different types of events. We found that users tended to spontaneously provide information about location during emergency events such as \textit{Abuse}, \textit{Break-In} and \textit{Robbery}. We also discovered that users tended to miss \textit{Attacker} information for \textit{Abuse} and \textit{Attack} events, and \textit{Target} information for \textit{Harassment} events.
Within our annotated sample, we also observed 34.8\% tips where users reported on behalf of victims or as witnesses. By conducting statistical analysis, it was also found that for different types of events, the amount of event information provided by users before dispatchers' involvement was significantly different. Meanwhile, we revealed that for events under \textit{HarassmentAbuse} category, the amount of information elicited by dispatchers differed significantly by event type.

%% file: 4_results_RQ2.tex
\subsection{How Do Dispatchers Interact with Users to Collect Information and Provide Emotional Support? (RQ2)}
After users provided the initial set of information through the submitted tip at the beginning of the conversation, dispatchers would proceed to collect more information from users by asking follow-up questions. Based on the proposed ontology, we annotated dispatchers' intent of questions asked to collect additional event argument information from users. We then analyze dispatchers' intent in order to understand the practice of dispatchers in the context of text-based safety reporting systems. 
In the following section, we first discuss dispatchers' behavior patterns when collecting information during conversations with varying depths. Second, we further investigated and summarized how dispatchers deployed different strategies to provide users with emotional support, then we identified the inconsistency in dispatchers' behavior in responding and providing emotional support. Finally, we identified several factors associated with dispatchers' decisions on whether or not to provide users with emotional support. 

\textbf{Varied Conversation Depth for Information Collection}
In order to further analyze the strategies used by dispatchers when asking questions, we conduct a sequential analysis of dispatchers' intent from the annotated sample data. 
The annotated data sample consists of 69 conversations. Since our analysis focuses on sequences where dispatchers asked questions, we excluded conversations where dispatchers were observed to have not asked any questions about event-related information, leaving us with a final sample of 52 conversations with 103 questions asked. 
The sample contains 13 \textit{Stolen} events, 9 \textit{Harass} events, 9 \textit{Abuse} events, 8 \textit{Attack} events, 6 \textit{Threaten} events, 4 \textit{Robbed} events, and 3 \textit{Break-In} events.

For dispatchers' questions for collecting information, we annotated question intents based on the type of event argument information each question aimed to collect. The event arguments we annotated include \textit{Location}, \textit{Attacker}, \textit{Attacker-Movement}, \textit{Target}, \textit{Weapon}, \textit{Start-Time}, \textit{End-Time}, and \textit{Target-Object}.
For example, with the dispatcher's utterance as "What is your address?", we annotate the question intent as asking for \textit{Location}.


To better quantify the number of interactions that happened between the user and dispatcher during one conversation, we define the process of the user answering the dispatcher's question as a question-answer turn. Based on this, we describe the length of a conversation using the total number of question-answer turns that took place during the conversation as conversation \textbf{depth}. 
Conversations with higher depth might indicate that more information needs to collect for the incidents. Since dispatchers might exhibit different behavior in terms of asking questions when more information is needed to be collected, we analyze how dispatchers used different ways and strategies to collect event-related information from users. 

\begin{figure}[h!]
  \centering
  \includegraphics[width=\linewidth]{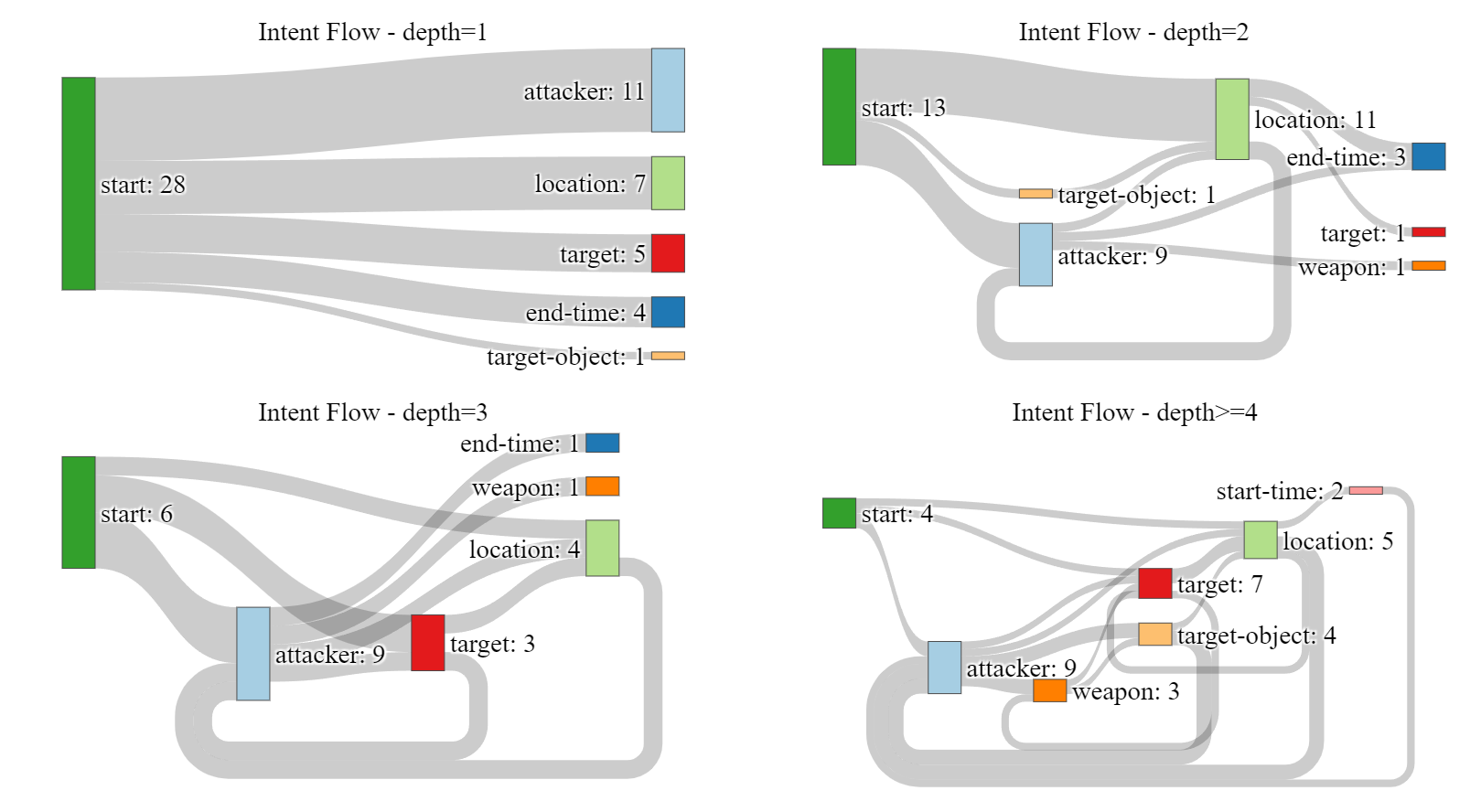}
  \caption{Sankey Diagram of Intent Sequences by Depth} 
  \label{fig:IntentSeq_by_depth}
\end{figure}

Among all questions, the top two most frequently asked types of event arguments by the dispatchers are \textit{Attacker} (69.2\%) and \textit{Location} (51.9\%). 
Similarly for both categories (\textit{HarassmentAbuse} and \textit{TheftLostItem}) of incidents, dispatchers tend to start with collecting \textit{Attacker} and \textit{Location} information. In the case of HarassmentAbuse, dispatchers are observed to start with collecting \textit{Target} information. 
As shown in Fig. \ref{fig:IntentSeq_by_depth}, for conversations ended with dispatchers' first replies, they tended to ask for information including \textit{Attacker}, \textit{Location}, \textit{Target} and \textit{Time}. This implied that reports often miss more information on these details.

Meanwhile, we also found that conversations with lower depth tend to have a more consistent flow, while the intent flow of dispatchers’ information-seeking utterances with higher depth tends to be more lacking in order or formality compared to ones with fewer turns.

A further Chi-squared test was performed to assess the relationship between the number of event information asked by dispatchers for different event types and conversation depth. There was a significant difference among conversations with different depths in the distribution of event information counts across event types ($X^2(110, N=3363) = 48.38, p = 0.007^{***}$). The results demonstrated that as conversational depth increased, so did the diversity of information dispatchers requested from users.
In the following sections, we further proceed to examine the detailed strategies used by dispatchers during conversations with users.


\textbf{Handling Reporters' Emotional Swings}
As is discussed in prior studies \cite{feldman2021calming, Observations_on_the_Display}, emergency line callers under circumstances such as medical and safety emergencies were observed to exhibit emotional displays that were characterized by a sharp increase or decrease in the level of emotion. These behaviors might include shouting, crying, speech distortion, etc.
Such emotional displays by reporters can present a challenge for the call-takers who need to maintain composure in order to provide effective assistance. However, in the context of a text-based reporting system, the reporters are not able to exhibit their emotional displays to the call-takers. 
In this section, we will discuss the results of case studies to answer two questions: 1) In the context of text-based reporting systems, how do reporters' emotional swings present themselves? and 2) How do call-takers handle reporters' emotional swings in order to provide effective assistance? 

As in the following example, 

\begin{quote}

\textbf{Incident category}: EmergencyMessage;
\textbf{Anonymous Tip}: No;

\vspace{0.2cm}

\textbf{User:} "I have really bad chills and I am having a hard time breathing."

\textbf{Dispatcher:} \textit{"what is your location and name?"}

\textbf{User:} "[LOCATION]"

\textbf{User:} "My name is [PERSON]"

\textbf{Dispatcher:} \textit{"[PERSON], what room number and would you like a ambulance?"}

\textbf{User:} "[LOCATION]"

\textbf{Dispatcher:} \textit{"[PERSON], I will send a officer to your location"}

\textbf{User:} "Thank you"

\textbf{Dispatcher:} \textit{"Officer [PERSON] is on his way"}

\textbf{Dispatcher:} \textit{"[PERSON], are you diabetic?"}

\textbf{Dispatcher:} \textit{"[PERSON], do you have a inhaler?"}

\textbf{User:} "No no"

\textbf{Dispatcher:} \textit{"ok, just take some deep breaths to slow down your breathing"}

\textbf{Dispatcher:} \textit{"has this happened before?"}

\textbf{User:} "I just got diagnosed with panic disorder"

\textbf{User:} "Yes it has"

\textbf{Dispatcher:} \textit{"ok Officer [PERSON] just arrived to the building"}

\textbf{User:} "I feel dizzy"

\textbf{Dispatcher:} \textit{"[PERSON], just take some slow breathes"}

\textbf{User:} "I'm so scared"

\textbf{Dispatcher:} \textit{"[PERSON], I'm with you and officer [PERSON] and officer [PERSON] will be right there i promise"}

\textbf{Dispatcher:} \textit{"[PERSON], I want you to continue to breathe slowly"}

\textbf{Dispatcher:} \textit{"[GPE] did you take any medication?"}

\textbf{User:} No

\textbf{Dispatcher:} \textit{"ok officer [PERSON] is in the elevator"}

\textbf{Dispatcher:} \textit{"He is out of the elevator you should see him in a minute"}

\textbf{User:} Ok

\textbf{Dispatcher:} \textit{"remember slow breaths"}

\textbf{User:} "I'm in my room"

\textbf{Dispatcher:} \textit{"ok he knows"}

\textbf{Dispatcher:} \textit{"Are you still dizzy?"}

\textbf{Dispatcher:} \textit{"[PERSON], you doing okay?"}

\textbf{Dispatcher:} \textit{"officers are at the location"}

\end{quote}

The user exhibited emotional swings, from "I have really bad chills and I am having a hard time breathing" to "I'm so scared". To calm the user from such emotional swings, the dispatcher reacted by 1) providing reassurance and support (e.g., "I'm with you and officer [PERSON] and officer [PERSON] will be right there I promise"), 2) asking questions to assess the situation (e.g., "Has this happened before?"), and 3) providing instructions to help the reporter (e.g., "remember slow breaths"). When using the text-based reporting system, the user displayed emotional traits directly through the use of words such as "scared", "dizzy", and "chills". 
Compared to  the phone call example \cite{feldman2021calming}, the text-based system might have lost some emotional information that can be helpful for dispatchers to understand the situation.
Meanwhile, due to the use of a text-only communication channel, less noise from the text was observed which can also help administrators to focus on the information provided by the user.

The dispatcher from the example also exhibited increased emotional awareness (e.g., from "[PERSON], I will send an officer to your location" to "I'm with you and officer [PERSON] and officer [PERSON] will be right there I promise") in response to the user's emotional swings. The emotional handling was conducted through text by the dispatcher in this example, by using the strategy of providing reassurance and support by frequently calling the user's name. The dispatcher was not able to use vocal cues or other nonverbal communication. 
This differs from the emotional swing handling under traditional call-taking situations, where the call-takers were observed to use verbal and nonverbal calming techniques such as speaking slowly and using a soft voice \cite{feldman2021calming}. 


\label{strategies}
By further examining other cases, we identified several other strategies that were also employed by administrators to handle users' emotional swings, as shown in Table \ref{Tab: emotional swing strategies}. In addition to \textit{Directives} and \textit{Reassurances}, which were methods that were also shown in prior work \cite{Observations_on_the_Display} to have been adopted by 911 dispatchers, we also revealed other strategies including \textit{Display of empathy}, \textit{Positive affirmation}, \textit{Providing additional support resources}, and \textit{Addressing users by first name}. 
We also found that dispatchers would not only provide \textit{Reassurances} to victims, but also to users that are not directly involved in the incident if reporting on behalf of the victim.

\begin{table}[!ht]
\resizebox{\linewidth}{!}{
\begin{tabular}{cl}
\hline
\textbf{Strategy used}                                                            & \multicolumn{1}{c}{\textbf{Sample replies from the dispatchers}}                                                                                                                                                                                                                                                                                                        \\ \hline
Directives                                                                        & \begin{tabular}[c]{@{}l@{}}“Ok, just take some deep breaths to slow down your breathing”\\ "Remain as quiet as possible. Turn your phone to silence to avoid being heard."\end{tabular}                                                                                                                                                           \\ \hline
\begin{tabular}[c]{@{}c@{}}Reassurances\\ (for victim)\end{tabular}      & \begin{tabular}[c]{@{}l@{}}“{[}PERSON{]} I'm with you”\\ “There is nothing to be sorry about. We are here to help.”\end{tabular}                                                                                                                                                                                                                  \\ \hline
\begin{tabular}[c]{@{}c@{}}Reassurances \\ (for witness)\end{tabular}             & “I am glad you are a good friend and let her talk, please know we are here for her and you.”                                                                                                                                                                                                                                                      \\ \hline
Display of empathy                                                                & \begin{tabular}[c]{@{}l@{}}“I am glad to hear he is currently at {[}FAC{]} receiving qualified assistance.”\\ “I am sorry to hear that happened to you.”\\ “sorry to hear that your bike was stolen”\end{tabular}                                                                                                                                 \\ \hline
Positive affirmation                                                              & “Thank you for being observant and reporting things that you feel are out of the ordinary.”                                                                                                                                                                                                                                                       \\ \hline
\begin{tabular}[c]{@{}c@{}}Providing additional \\ support resources\end{tabular} & \begin{tabular}[c]{@{}l@{}}“There is also a hotline she can call, I am looking for the number. \\ Either one of the hospitals is confidential. I also believe {[}ORG{]} is too.”\\ “Speaking to them (Officers) or having a conversation with them \\ doesn't mean she will be forced to seek help, sometimes just an ear to listen”\end{tabular} \\ \hline
\begin{tabular}[c]{@{}c@{}}Addressing user \\ by first name\end{tabular}          & “{[}PERSON{]}, just take some slow breathes”                                                                                        \\ \hline
\end{tabular}
}
\caption{Example cases for strategies used by dispatchers to handle users' emotional swings}
\label{Tab: emotional swing strategies}
\vspace{-0.5cm}
\end{table}

These results suggest that, when facing with reporters' emotional swings during the text-based reporting process, the dispatchers need to use different emotional handling strategies, compared with the phone call context. They need to be more direct and explicit in their emotional handling in order to provide effective assistance to the reporters.


\textbf{Administrators' Inconsistency in Response}
We used the complete dataset for this part of the analysis. We quantified responsiveness by calculating the average delay duration before administrators' every response. More specifically, we compute the average time elapsed per conversation after each message sent by the user and before the administrator's following utterance as the measure for the delay in response. 
We also found that administrators' responsiveness differs significantly across different event category using ANOVA ($F(5, 1951) = [3.49], p = 0.003^{***}$). 
Tukey post hoc test revealed that administrators respond to users' message slower under \textit{TheftLostItem} categories than \textit{DrugsAlcohol} ($p = 0.02^{**}, 95\% \  C.I. = [3.82, 69.31]$), \textit{EmergencyMessage} ($p = 0.006^{***}, 95\% \  C.I. = [7.05, 71.77]$) and \textit{SuspiciousActivity} ($p = 0.002^{***}, 95\% \  C.I. = [10.61, 71.51]$). 
We also identified that administrators' responsiveness differs significantly across different organizations by conducting an ANOVA analysis ($F(110, 3252) = [2.57], p < 0.001^{***}$). 

We then examined the change in administrators' tone when collecting incident information from users.
We found that even for the same organization, administrators sometimes used different tones when chatting with users. As an example, from an organization (that implemented the system for 5.3 years) we found during one \textit{HarrassmentAbuse} incident: 

\begin{quote}

\textbf{Incident category}: HarassmentAbuse;
\textbf{Event Type}: Harass;
\textbf{Anonymous Tip}: Yes;

\vspace{0.2cm}

...... 

\textbf{Dispatcher:} \textit{"Ok we have notified officers and will check area, was your friend injured? Does she need medical attention?"}

\textbf{User} "No! She is fine. She just said everytime she tried exiting the conversation he was grab her shoulder and continue the convo"

\textbf{Dispatcher:} \textit{"I'm very glad your friend is fine! We have officers checking the area for the male subject. Would she like to meet with officers?"}

......

\end{quote}

Where the dispatcher actively expressed emotional support and confirmation to the user. Such use of emotional support and empathy exhibition was not observed for most other conversations for the same organization under the same incident category, as in the following conversation where another \textit{HarrassmentAbuse} incident happened:

\begin{quote}

\textbf{Incident category}: HarassmentAbuse;
\textbf{Event Type}: Abuse;
\textbf{Anonymous Tip}: Yes;

\vspace{0.2cm}
......

\textbf{Dispatcher:} \textit{"Thank you. Officers are at the location. Can you describe the male that was pushing the female?"}

\textbf{User} "[SUSPECT DESCRIPTION]"

\textbf{Dispatcher:} \textit{"Thank you."}

\textbf{User} "Can you let me know if they were able to talk to them? Very concerned for the girl"

\textbf{Dispatcher:} \textit{"Officers at(are) there now."}

\textbf{User} "Were they able to find the couple?"

\textbf{Dispatcher:} \textit{"Yes, Officers are with them now."}

......

\end{quote}

From the conversations, we found that many administrators tend to repetitively say thank you and confirm officers' departure or arrival, while no emotional support or confirmation was provided to users. This might be because responses are often based on given guidelines or playbooks, where emotional support is often not covered as part of the standard procedure or practice. As a result, responses can often be standardized and appear to lack empathy toward community members. 

For different organizations, we also observed an organizational discrepancy in providing emotional support. Based on the strategies we discovered in Table \ref{Tab: emotional swing strategies}, we sampled conversations from the top-3 organizations with the most tips and annotated the strategies used by their dispatchers.

For the cases we reviewed, we found that different organizations varied in both tones of speech and strategies in providing emotional support. 
As shown in Table \ref{Tab:Grid_Table_Emotion_strats}, Organization A was observed to have mainly adopted the method of expressing empathy (e.g., "I’m so sorry for what happened to you"), providing positive affirmation (e.g., "Thank you for being vigilant and using LiveSafe"), and addressing the user by the first name (e.g., "[NAME] we have to send an officer to take the report in person"). 
Organization B adopted the strategy of providing reassurances (e.g., "We don’t either, this has been a very hard time for so many people") that was not observed by organization A or C. Meanwhile, expressions of empathy and positive affirmation were also observed.
In contrast, organization C took on a more formal tone of speech. We observed that the dispatcher used directives in order to calm the user and collect information (e.g., "I need to know where you are and what you drive to send someone to you"). Reassurance and display of empathy were also found to be provided by the dispatcher even when the reporter was not the victim (e.g., "Thank you for this additional information. I am glad to hear he is currently at [FAC] receiving qualified assistance"). 


\begin{table}[ht!]
\resizebox{\linewidth}{!}{
\begin{tabular}{clllllll}
\hline
\textbf{Organization}               &  \multicolumn{7}{c}{\textbf{Emotional Support Strategies}}          \\
\hline
\multicolumn{1}{l}{} & \rot{Directives}                                                                 & \rot{Reassurances}                                                              & \rot{\parbox{2.5cm}{Reassurances \\ (for witness)}}                                                     & \rot{Display of empathy}                                                     & \rot{Positive affirmation}                                                      & \rot{\parbox{2.8cm}{Providing \\ additional resources}}                                                     & \rot{\parbox{2.5cm}{Addressing user \\ by first name}}                                                     \\ \hline
Org. A               &                                            &                                            &                                                                                            & \checkmark                                                                                                     & \checkmark                                                               &                                                                                  & \checkmark                                                                        \\
Org. B               &                                            & \checkmark                      &                                                                                            & \checkmark                                                                                                     & \checkmark                                                               &                                                                                                       &                                                                                              \\
Org. C               & \checkmark                      &                                            & \checkmark                                                                      & \checkmark                                                                                                     &                                                                                     & \checkmark                                                                                &                                                                                              \\ \hline
\multicolumn{8}{c}{\textbf{Quotes from each organization from multiple conversations}}                                                                                                                                                                                                                                                                                                                                                                                                                                                                                                                                                       \\ \hline
Org. A               & 
\multicolumn{7}{l}{\begin{tabular}[c]{@{}m{12cm}@{}}
\textbullet{} Thank you for being vigilant and using LiveSafe. \\
\textbullet{} Ok. Well I'm glad to hear you are not hurt. \\
\textbullet{} {[}NAME{]} we have to send an officer to take the report in person.\\ 
\textbullet{} Of course, and I'm so sorry for what happened to you.\end{tabular}}      \\ \hline
Org. B               & \multicolumn{7}{l}{\begin{tabular}[c]{@{}m{11.5cm}@{}}
\textbullet{}  We don't either, this has been a very hard time for so many people. \\ 
\textbullet{}  Thank you so much for giving us this information. 
Also if you need anything don't ever hesitate to reach out to us, we are always here 24/7. \\
\textbullet{}  Glad you and others on the road are not hurt, and hopefully we can apprehend them before someone is hurt. Our officers have the information.\end{tabular}}                                                                                                                                                                                                                                                                                                 \\ \hline
Org. C               & \multicolumn{7}{l}{\begin{tabular}[c]{@{}m{12cm}@{}}
\textbullet{} Okay. I'm very sorry this happened. What you can do is utilize our online reporting system and start a report for {[}ORG{]}. You can find that at {[}WEBSITE{]} and it will walk you through the steps of filing that report.\\ 
\textbullet{} Thank you for this additional information. I am glad to hear he is currently at {[}FAC{]} receiving qualified assistance.\\ 
\textbullet{} I need to know where you are and what you drive to send someone to you. If I don't know your location, I'm sorry I can't send anyone to you.\end{tabular}} \\ \hline
\end{tabular}
}
\caption{Example cases of emotional support strategies used by different organizations}
\label{Tab:Grid_Table_Emotion_strats}
\vspace{-0.5cm}
\end{table}

\textbf{When will Administrators Provide Emotional Support?}
Our previous analysis revealed the inconsistency in administrators' decisions on whether or not to provide emotional support to users. Additional analyses were carried out to better understand the potential factors influencing these decisions in order to examine dispatchers' behavior in terms of offering emotional support. We utilized a T5-based emotion classifier as described in \ref{sec: method_t5} to detect utterances where administrators provided emotional support.



\begin{figure*}[!ht]
\centering

\begin{subfigure}{.48\linewidth}
  \centering
  \includegraphics[width=\linewidth]{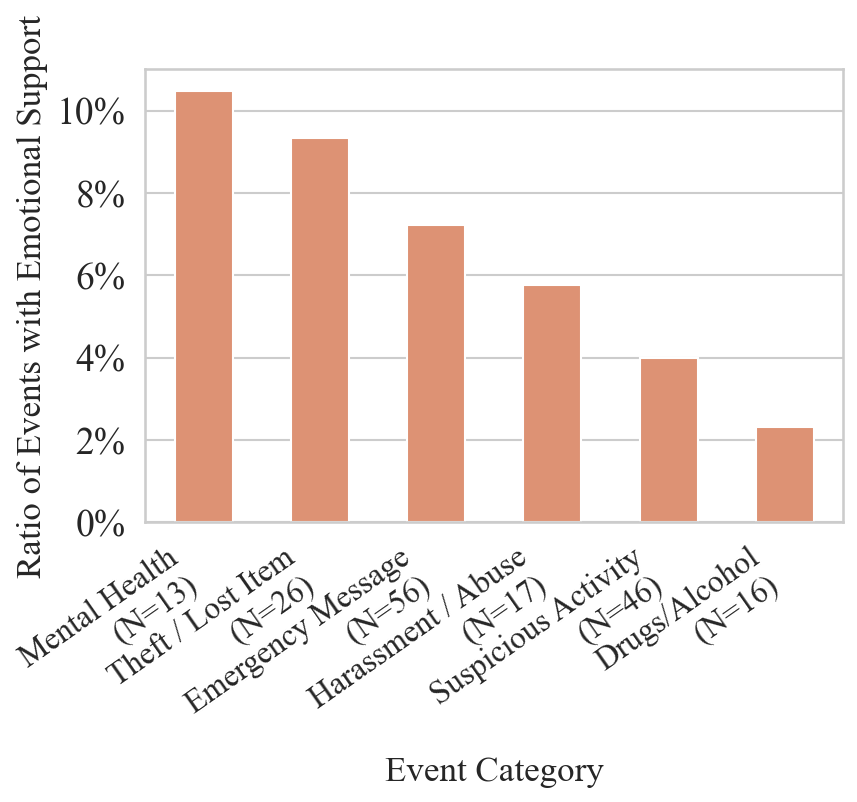}
\caption{Ratio of tips where dispatchers provided emotional support by category}
\label{fig:emotional_support_ratio}
\end{subfigure}%
\hspace{0.3cm}
\begin{subfigure}{.48\linewidth}
\centering
  \includegraphics[width=\linewidth]{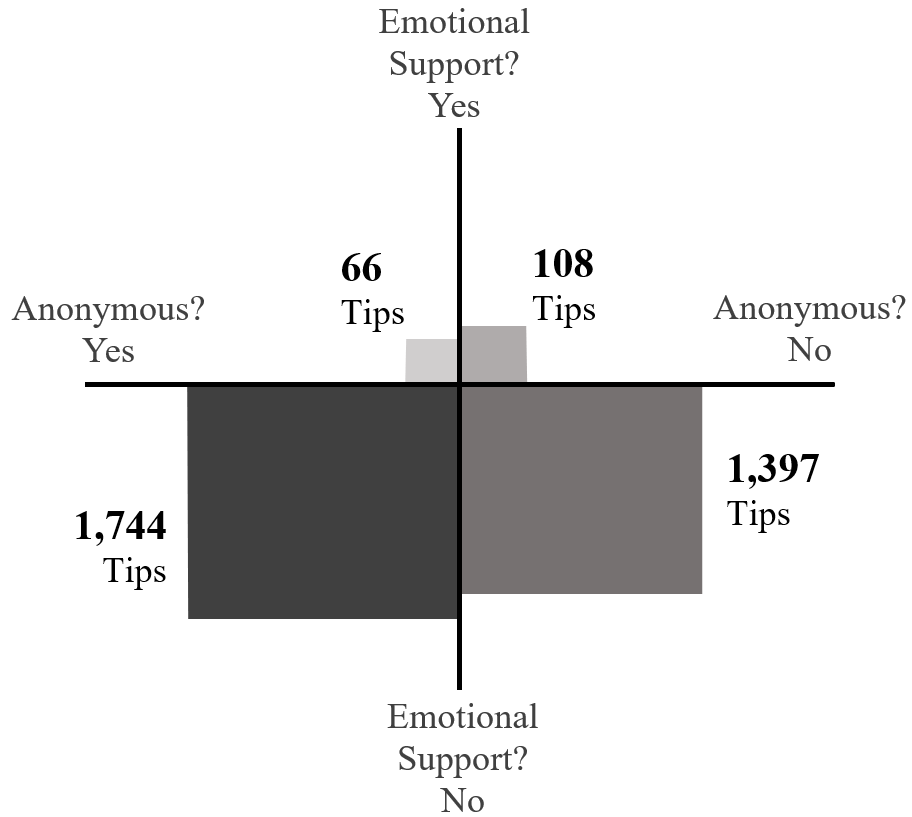}
\caption{Fourfold plot of tips distribution by anonymity and emotional support}
\label{fig:emotional_support_quad}
\end{subfigure}
\caption{Description of Tips with Detected Emotional Support}
\end{figure*}

As shown in Fig. \ref{fig:emotional_support_ratio}, \textit{Mental Health} tips had the highest ratio of events where administrators provided emotional support. We also identified a higher proportion of such tips under \textit{Theft Lost Item} category. By examining individual cases, we found that during most of these \textit{Theft Lost Item} cases administrators were observed to have expressed empathy for users' loss (e.g., "I'm sorry that happened", "I am sorry this has happened"). We also found that when compared to tips that were submitted anonymously, tips submitted along with identifiable information received more emotional support from administrators, as shown in Fig. \ref{fig:emotional_support_quad}.

In order to further understand under what circumstances emotional support was provided by dispatchers, we conducted a logistic regression to analyze the relationship between event categories, reporter role (if reporting on behalf), anonymity, Organization's age since implementation, and if a tip is provided with emotional support. 
We also controlled variables including \textit{time of day} and \textit{season}, since night shifts and student volumes during different semesters may also affect dispatchers' incident handling behavior.

\begin{table}[t!]
\resizebox{\linewidth}{!}{
\begin{tabular}{p{7cm}p{1.3cm}p{2.5cm}p{1.3cm}p{2.5cm}}
\hline
\multicolumn{1}{c}{}                       & \multicolumn{2}{l}{Model 1} & \multicolumn{2}{l}{\begin{tabular}[l]{@{}l@{}}Model 2 \\ (w/ interaction term)\end{tabular}} \\ \hline
\multicolumn{1}{c}{}                       & Coeff.             & (S.E.) & Coeff.                                             & (S.E.)                                  \\ \hline
Time of day                                &                    &        &                                                    &                                         \\
\hspace{3mm} 4 a.m. - 8 a.m.                            & REF                &        &                                                    &                                         \\
\hspace{3mm} 8 a.m. - 12 p.m.                           & -0.06              & (0.44) & -0.06                                              & (0.44)                                  \\
\hspace{3mm} 12 a.m. - 4 p.m.                           & -0.08              & (0.41) & -0.09                                              & (0.41)                                  \\
\hspace{3mm} 4 a.m. - 8 p.m.                            & 0.02               & (0.41) & 0.02                                               & (0.41)                                  \\
\hspace{3mm} 8 a.m. - 12 a.m.                           & 0.19               & (0.40) & 0.19                                               & (0.40)                                  \\
\hspace{3mm} 12 a.m. - 4 a.m.                           & 0.05               & (0.42) & 0.05                                               & (0.42)                                  \\
Season                                     &                    &        &                                                    &                                         \\
\hspace{3mm} Spring                                     & REF                &        &                                                    &                                         \\
\hspace{3mm} Summer                                     & -0.05              & (0.28) & -0.05                                              & (0.28)                                  \\
\hspace{3mm} Fall                                       & 0.04               & (0.21) & 0.04                                               & (0.21)                                  \\
\hspace{3mm} Winter                                     & 0.19               & (0.20) & 0.19                                               & (0.20)                                  \\
Org. age since implementation              & -0.01              & (0.07) & -0.01                                              & (0.07)                                  \\
\textbf{Event category}                             &                    &        &                                                    &                                         \\
\hspace{3mm} Suspicious activity                        & REF                &        &                                                    &                                         \\
\hspace{3mm} \textbf{Drugs/Alcohol}                              & \textbf{-0.51*}              & (0.30) & \textbf{-0.51*}                                              & (0.30)                                  \\
\hspace{3mm} \textbf{Emergency message }                         & \textbf{0.48**}     & (0.21) & \textbf{0.48**}                                     & (0.21)                                  \\
\hspace{3mm} Harassment/Abuse                           & 0.45               & (0.29) & 0.45                                               & (0.29)                                  \\
\hspace{3mm} \textbf{Mental Health}                              & \textbf{1.12***}   & (0.33) & \textbf{1.12***}                                   & (0.33)                                  \\
\hspace{3mm} \textbf{Theft/Lost item}                            & \textbf{0.70***}    & (0.27) & \textbf{0.70***}                                    & (0.27)                                  \\
\textbf{Reporting on behalf/as witness}             & \textbf{-1.29**}    & (0.51) & \textbf{-1.44**}                                    & (0.72)                                  \\
\textbf{Anonymity}                                  & \textbf{-0.48***}   & (0.18) & \textbf{-0.48***}                                   & (0.18)                                  \\
Reporting on behalf/as witness $\times$ Anonymity &                    &        & 0.33                                               & (1.03)                                  \\ \hline
{\scriptsize * p < 0.1; ** p < 0.05; *** p < 0.01 (two-tail); S.E. clustered by organization}
\end{tabular}
}
\caption{Results of logistic regression investigating the factors that are associated with emotional support provided for incident tips (N = 3315); Reporting on behalf/as witness means whether the user reporting the incident was the victim or not.}
\label{Tab: emotional_support_Logistic}
\vspace{-10pt}
\end{table}

As shown in Table \ref{Tab: emotional_support_Logistic}, we found that tips with category \textit{EmergencyMessage}, \textit{MentalHealth} and \textit{TheftLostItem} were more likely to receive emotional support comparing to \textit{SuspiciousActivity}, among which \textit{MentalHealth} had the highest odds ratio. 
A negative significant association was also found between if a user was reporting on behalf of others and the odds of the tip being provided with emotional support, which means users reporting on behalf of others were less likely to receive emotional support. 
Similarly, we also found that users reporting tips anonymously were less likely to receive emotional support. However, we did not identify any association between emotional support and the age of the organization since the LiveSafe system was implemented. 

\hspace{0.01mm}

\textbf{Summary}: By analyzing how dispatchers responded to users' reports, we first revealed the strategies dispatchers used to handle users' emotional swings,
we found that several strategies were used to calm users' emotional swings during conversations. Users were observed to be able to express emotional displays through text. Strategies including \textit{Display of empathy}, \textit{Positive affirmation}, etc were adopted during conversation with users. 
By conducting a logistic regression analysis, we revealed the factors associated with if a tip was provided with emotional support. A significant association was identified between the odds of a tip being provided with emotional support and factors including tip category, reporting on behalf, and user anonymity. 

%% file: 4_results_RQ3.tex
\subsection{How Do Users Engage in Dispatchers’ Follow-ups? (RQ3)}

In this section, we conduct analyses to understand the factors associated with users' responsiveness toward questions.


\textbf{Higher User Engagement When Provided with Emotional Support} 
In order to understand how users respond, similarly to dispatcher responsiveness, we calculated the average time elapsed per conversation after a message sent by the organization and before the user's reply, as the measurement of individual responsiveness. 
To further understand what factors are associated with users' responsiveness, we conducted a linear regression analysis to assess the relation between users' average time elapsed per message and factors including event categories, emotional support, anonymity. We also included administrators' average time elapsed per message to measure their responsiveness. 

As is shown in Table \ref{Tab: elapsed_user_Linear}, users who received emotional support showed higher responsiveness: that is, there was significantly less delay in their responses. When administrators responded slower, we similarly found that users showed a significantly higher delay in their responses. However, no significant association was found between users' anonymity and responsiveness. 
The findings revealed that, in addition to the urgency of the reported incident, the quality of service offered by dispatchers (i.e., emotional support and dispatcher responsiveness) may have an impact on users' responsiveness during the reporting process.

\begin{table}[ht!]
\resizebox{\linewidth}{!}{
\begin{tabular}{p{7cm}p{1.3cm}p{2.5cm}p{1.3cm}p{2.5cm}}
\hline
\multicolumn{1}{c}{}                       & \multicolumn{2}{l}{Model 1} & \multicolumn{2}{l}{\begin{tabular}[l]{@{}l@{}}Model 2 \\ (w/ interaction term)\end{tabular}} \\ \hline
                                            & Coeff.            & (S.E.)  & Coeff.                                             & (S.E.)                                  \\ \hline
\textbf{Event category}                              &                   &         &                                                    &                                         \\
\hspace{3mm} Suspicious activity                         & REF               &         &                                                    &                                         \\
\hspace{3mm} Drugs/Alcohol                               & 0.26              & (0.20)  & 0.26                                               & (0.20)                                  \\
\hspace{3mm} \textbf{Emergency message}                           & \textbf{-0.38*}   & (0.20)  & \textbf{-0.38*}                                    & (0.20)                                  \\
\hspace{3mm} Harassment/Abuse                            & 0.27              & (0.27)  & 0.27                                               & (0.27)                                  \\
\hspace{3mm} Mental Health                               & -0.28             & (0.39)  & -0.28                                              & (0.39)                                  \\
\hspace{3mm} Theft/Lost item                             & 0.43              & (0.28)  & 0.43                                               & (0.28)                                  \\
\textbf{Dispatcher’s response delay per message} & \textbf{0.18***}  & (0.03)  & \textbf{0.18***}                                   & (0.03)                                  \\
\textbf{w/ Emotional support}               & \textbf{-0.75**}  & (0.32)    & \textbf{-0.72*}                                    & (0.41)                                  \\
Anonymity                          & 0.18              & (0.16)  & 0.18                                               & (0.16)                                  \\
w/ Emotional support $\times$ Anonymity            &                   &         & -0.07                                              & (0.66)                                  \\ \hline
{\scriptsize * p < 0.1; ** p < 0.05; *** p < 0.01 (two-tail); S.E. clustered by organization}
\end{tabular}
}
\caption{Results of Linear Regression Investigating the Factors that are Associated with Users' Delay in Response for Incident Tips (N = 3315);}
\label{Tab: elapsed_user_Linear}
\vspace{-10pt}
\end{table}


\textbf{Questions Leading to Incomplete Conversations} 
Within the chat logs, we observed cases where the conversation ended with an administrator question that did not receive a user's response. To further examine these incomplete conversations, we extracted all conversations ending with a dispatcher message containing a question mark from the complete dataset. Then, we manually reviewed all the extracted cases and categorized them by question intent. 
By doing this, we aimed to identify all conversations where users did not respond to an asked question. As a result, we collected a total of 94 incomplete conversations.
We then analyzed the question types where users did not respond. 

\begin{figure}[ht!]
  \centering
  \includegraphics[width=\linewidth]{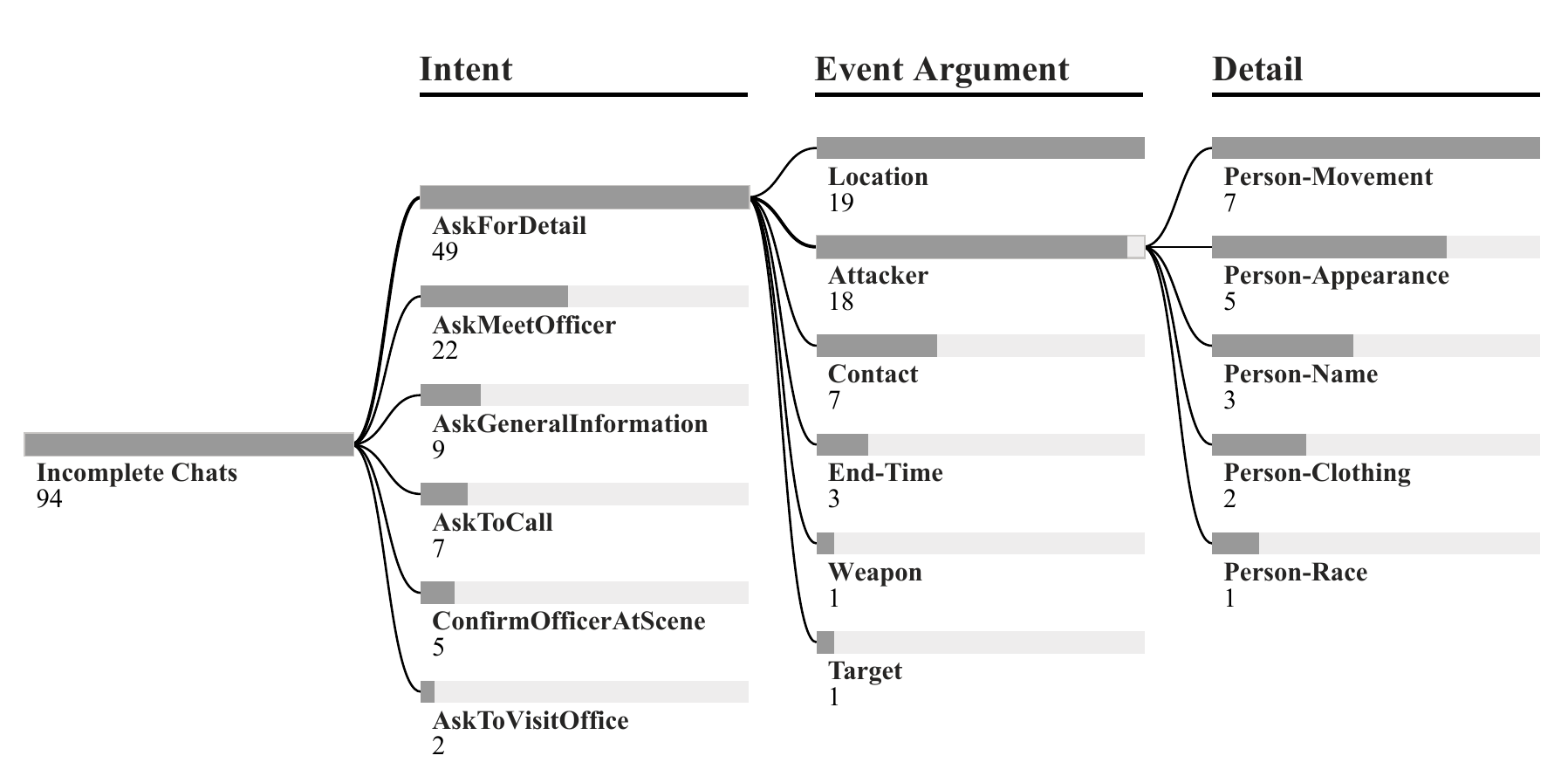}
  \caption{Dispatchers' question intents leading to incomplete conversations} 
  \label{fig:Incomplete_Chats}
\end{figure}

As shown in Fig. \ref{fig:Incomplete_Chats}, there were 31 conversations (33.0\%) where the user did not respond after the dispatcher requested to communicate using other traditional channels. This included asking the user to meet an officer in person, call the safety phone line, and visit the safety department. In these situations, it is possible that individuals continued the conversation through these channels, but without specific data in the logs we are unable to verify any follow-on activity.
Additionally, 49 conversations (52.1\%) ended with the dispatcher asking the user for specific event-related information, such as information related to \textit{Location}, \textit{Attacker}, and the user's \textit{Contact} information, leading us to infer that users may end conversations when asked questions that would compromise their anonymity.
More specifically for information related to \textit{Attacker}, we found that users sometimes did not respond to questions about the attacker's description details, such as \textit{Movement}, \textit{Appearance}, \textit{Name}, etc. 
We also noted that during 9 conversations (9.6\%) the conversation ended with the dispatcher asking the user for other general information (e.g., "Is there anything other information you could provide us with?"). This type of question might be perceived by the user as an alternative way to end the conversation. 

\textbf{Users Having Lower Responsiveness than Dispatchers} 
By conducting ANOVA over all conversations under categories of \textit{HarrassmentAbuse} (N=299) and \textit{TheftLostItem} (N=280), it was found that users (M=2.6 minutes, SD=5.0 minutes) had significantly longer time elapsed before response than administrators (M=1.8 minutes, SD=2.9 minutes)  ($F(1, 1131) = [12.44], p < 0.001^{***}$). In particular, the variation in results suggest that users had different level of responsiveness than dispatchers.

This corresponds to the fact that users sometimes report incidents on the move, as in the following example:

\begin{quote}

\textbf{Incident category}: HarassmentAbuse;
\textbf{Event Type}: Harassment;
\textbf{Anonymous Tip}: No;
\vspace{0.2cm}

\textbf{User} "there are boys from [PERSON] catcalling and surrounding girls on scholars"

\textbf{Dispatcher:} \textit{"sending an officer now"}

\textbf{Dispatcher:} \textit{"he's on his way"}

\textbf{Dispatcher:} \textit{"Where is this happening? Officer said no one is out there."}

\textbf{User} "we ran back to [LOCATION]. i think they left. i just wanted to notify y'all. that there was guys walking around cat calling and surrounding girls"

\textbf{User} "i didn't know if they were still there"

\textbf{Dispatcher:} \textit{"ok, I'll let officer know"}

\end{quote}

In the case above, there existed a gap of 8 minutes between the first and second message of the user. With the aid of the text-based reporting system, the user was able to report the incident on the move and provide the dispatcher with additional follow-up information, which would have not been possible using other forms of communication, such as a phone call.

\hspace{0.01mm}

\textbf{Summary}: 
To summarize, we conducted a linear regression analysis and revealed that both  emotional support and dispatchers' responsiveness are positively associated with users' responsiveness.
Through examining the incomplete conversations where the user did not respond to the dispatcher's last question, we discovered findings that implied users' aversion to responding to questions that could potentially compromise their anonymity. 
We also found that users responded more slowly to dispatchers’ questions than dispatchers responded to users. Cases were identified where users temporally paused the conversation but were able to resume and provide more information. 

%% file: 5_discussion.tex
\section{Discussion}
In this section, we discuss the findings of this study and their connections to prior work and implications for future designs in the field of community safety risk reporting systems. 

\subsection{\textit{Hidden Facts}: Stakeholders'  Interactions through Text-based Reporting Systems}

We first focus on discussing what new understanding of user-dispatcher interaction in a text-based reporting system has been revealed by the results of this study.

\textbf{Misaligned Perceptions on Risk Reporting Using Text-based Channels (RQ1)} 
Previous work identified that users are more likely to report an incident directly by dialing 911 when encountering emergencies and life-threatening incidents than using other channels of communication \cite{hughes2012evolving}.
Examples of emergency dispatch center training manuals also stated that callers under emergency situations should be routed to emergency lines immediately, reinforcing the role of these systems for high-severity incidents \cite{alameda_police_2020}. 
Our findings suggested that even under emergency situations, some users still choose to report through text-based systems. This is aligned with findings from recent studies about newly-emerged ICTs and their application for crisis management \cite{schraagen2011human} and incident reporting \cite{tabassum2022context}. 
Online text-based reporting may be perceived by users as more private and anonymous \cite{alarid2008citizens}, which might encourage users to report.
In addition, some users might use the text-based system if they are unable to speak, such as a user experiencing difficulty breathing or needing to report quietly. This type of situation has also been identified as a significant challenge for traditional call-taking systems \cite{neusteter2019911}. 
Our findings showed that under such situations, dispatchers were able to effectively communicate and offer both physical and mental support to users through the text-based communication channel. 

Our findings from RQ1 also showed that although users tended to provide similar types of event information upfront (e.g., \textit{attacker}, \textit{victim}, \textit{location}, and \textit{time}), the types of information asked and elicited by dispatchers often differed across different event types. Meanwhile, the amount of information reported by users was also found to differ across event types. 
This suggests a difference between the information that is important from users' perception and that from safety organizations. This is also referred to as the "discrepancy between the perceived priority of concerns" by a previous study by Whalen et al.\cite{Observations_on_the_Display}. 
The implications of this finding are twofold: First, more information should be offered to align users' and dispatchers' perceptions on the details that are most important for responding to events. This can be accomplished by providing a well-designed event ontology with explicit explanations to help users and dispatchers communicate more effectively. Second, this conclusion emphasizes the significance of flexibility when developing new ICTs for the safety reporting area in order to capture information with multiple modalities for various sorts of occurrences. A flexible and adjustable event reporting system can help to bridge the gap between what information is most valuable to each stakeholder.



\textbf{Inconsistency between Dispatchers' Responses (RQ2)}
According to prior works \cite{Observations_on_the_Display, feldman2021calming, paoletti2012operators}, 911 emergency line call-takers often have to handle callers' strong emotional displays in order to offer assistance and gather information. 
According to our RQ2 findings, we also observed similar behavior from users of the LiveSafe text-based reporting system. Using text, users were still observed to express emotional swings. We also identified several strategies used by administrators in response to users' emotions. 
In addition to strategies discussed in prior works\cite{Observations_on_the_Display, feldman2021calming}, we identified additional strategies specifically pertaining to expressions of emotional support from dispatchers. This resonates with previous works about the benefit of emotional support for crime victims \cite{DakofVictimsPerceptions}. 

However, we found that providing emotional support was not a standardized procedure consistently adopted by dispatchers under all circumstances. 
Our findings from RQ2 showed that not only does an organizational discrepancy exist in the practice of providing emotional support, but dispatchers' decisions on whether to provide emotional support were also found to be affected by profiles of users (i.e., anonymity and if reporting on behalf). Additionally, the responsiveness of dispatchers was seen to vary among event types.
This shed light on how responsiveness can be affected by administrators' perception of risk. Miscommunication and other undesirable outcomes might emerge from a mismatch in perceptions of risk on each side of the conversation \cite{garcia2015something}. ICTs can be utilized by agencies to improve the consistency in safety services to improve community trust \cite{SheenaExaminingTechnologyThat}.



\textbf{The Myths of Emotional Support (RQ3)} 
Prior research on text-based reporting systems has revealed the mismatch between what the safety departments expected to receive and what their community members actually reported \cite{ming2021examining}. This is due to a misalignment between the agency's restricted number of tip categories and how these categories are perceived by users. Users may be hesitant to report incidents as a result of this mismatch, lowering the level of service provided by safety agencies.
Previous research has emphasized the significance of service quality to emergency dispatch operations \cite{clawson2018litigation}. In this study, we discovered two additional organizational factors related to service quality that are associated specifically with users' responsiveness in reporting based on RQ3 findings: organizational emotional support and responsiveness, with users' responsiveness being positively associated with provided emotional support and administrator responsiveness.
The findings imply that administrators' behavior has a significant impact on community members' reporting responsiveness. This is consistent with other studies where it was discovered that users' reporting decisions were influenced by their level of trust in the effectiveness of the police \cite{goudriaan2006neighbourhood}. 

In the utility-cost model of users' reporting behavior, the decision of reporting is driven by users' estimation of benefits if a report is made and the cost of doing so. Prior works have suggested that victims who have low confidence in the efficacy of safety organizations may not value the benefits of reporting as highly \cite{baumer2002neighborhood, anderson2000code}. 
As a result, our study added empirical support to the idea that the responsiveness of users' reporting behavior may be influenced by safety agencies' service quality. To further quantify the process of users' reporting behavior, we propose to consider the reporting behavior as a sequential decision-making process, where methods such as the nested logit model may be adopted to better model and understand users' behavior \cite{lemp2010continuous}.
Another implication of our findings is that safety agencies can improve the quality of their services by providing emotional support and improving responsiveness to encourage users to report safety concerns and events.


\subsection{Design Implications}

\textcolor{blue}{In this section, we discuss the design implications of our findings from two main perspectives: one is regarding improving the existing  features of text-based reporting systems (in Section \ref{general}); and the other is about creating an AI-enabled conversational agent (CA) that can mediate user-dispatcher interaction (in Section \ref{ca}).}




\subsubsection{Improving the Anonymous and Text-based Interactions between Two Stakeholders}\label{general}
\textcolor{blue}{As shown in Fig.~\ref{fig:livesafeInterface}, as a text-based reporting system, \textit{LiveSafe} shared two major features with many mobile safety reporting apps: 1) users can send anonymous tips and 2) users and dispatchers have chat in text  asynchronously. } 

\textcolor{blue}{First, anonymity has been identified as an essential design option in many existing reporting systems \cite{ming2021examining, o2006guarantee, campbell2019risking}.
Though anonymity could impact users' reporting behavior \cite{ming2021examining}, e.g., reducing users' concerns and improving the quantity of reporting,  our findings also showed that anonymous reports were less likely to receive dispatchers' emotional support. Future reporting systems could detect whether emotional support is missing in dispatchers' messages and remind dispatchers to provide such service. If it's not an anonymous tip, then \textit{addressing users by their first names}  strategy (as shown in \ref{strategies}) that addressing users' names could be recommended as one option; otherwise, strategies such as \textit{reassurance} and \textit{display of empathy} (as shown in \ref{strategies}) could be offered for dispatchers to consider. This design can help dispatchers have varied strategies to provide emotional support.}

\textcolor{blue}{Second, it is intuitive that text-based chats gave users a lot of flexibility. For example, our findings (RQ3) showed that the asynchronous chats allowed users to confirm for the resolution of the incident, and their location changes during a prolonged reporting period. However, this posed a lot of challenges for the dispatchers, who had to constantly ``pull'' information from the users by asking if the users could respond to the dispatchers' questions. Unlike synchronous chats (e.g.,\cite{suler2011psychology}),  users often had delayed confirmations using asynchronous texting (RQ3), which could negatively impact the efficiency of the communication. Future systems may allow dispatchers and users to set up default communication mechanisms for handling ``delayed'' responses. For example, depending on the urgency of the incidents, time-based pull requests could be automatically sent to the users periodically, and the period and frequency could be customized by the dispatchers and users. The system-supported requests could be location-based as well, e.g., when users' locations are changed, reminders are prompted to users to follow up with the dispatchers. By doing so, the system can support more timely communication with a combination of pull- and push-based communication. 
}

\subsubsection{Designing Conversational Agents for High-Quality Reporting}\label{ca}
Inspired by the conversational nature of text-based reporting systems, we identified the potential of incorporating an AI-enabled conversational agent (CA) into the user-dispatcher interaction during incident reporting. 
We focus on discussing the implications of our findings for introducing a \textit{\textbf{Human-CA Collaborative design}} that can augment the interaction between users and dispatchers in the context of text-based risk reporting systems. Specifically, we discuss the potential of using CAs to bootstrap responses for dispatchers from four perspectives: 
1) efficient incident information extraction; 
2) emotional intelligence; 
and 3) awareness of broader contexts.

\textbf{Automating Safety Incident Information Collection} 
Our findings revealed that during a reporting conversation initiated by users with dispatchers, the common interaction process is that the dispatcher asks questions to collect information from the user’s response. We also discovered that dispatchers are inconsistent in 1) their responsiveness, 2) strategies used, and 3) whether or not they provide emotional support during this procedure.
Recent work explored the feasibility of using CA to help increase the effectiveness and efficiency of collecting evidential statements from both victims and witnesses \cite{minhas2022protecting}. Our findings also contributed to the growing body of evidence that a conversational agent tool can be created to increase communication consistency and efficiency by facilitating interactions between users and dispatchers~\cite{zheng2022ux}. The CA might improve service quality by supporting dispatchers in producing better reply messages, which would increase users' responsiveness. It could also directly increase reporting efficiency by automating the process of gathering essential event argument information. 
However, evidence collection systems proposed by prior studies \cite{minhas2022protecting, ku2008crime} either rely on rule-based methods or require a large amount of manually labeled training data. 
Future studies can adopt methods from recent NLP works from the domain of dialogue state tracking \cite{wu2019transferable, gao2019dialog, lee2021dialogue} and event argument extraction \cite{du2020event, lyu2021zero}. We also suggest future system design practitioners consider using pre-trained language models for more effective and robust event information extraction, by increasing the model's ability in domain adaptation and commonsense reasoning \cite{yang2019exploring}.

From our findings of RQ1, we observed several cases where the user was reporting on behalf of another person (the victim). Such reporting behaviors are often seen in situations where victims were unable or willing to report for themselves. This has been pointed out by previous studies in the context of crimes and safety incidents such as harassment \cite{matias2015reporting} and sexual abuse \cite{allnock2019snitches}. As an implication of this finding, we suggest that the CA should also be enabled to collect information about users' role in the incident, i.e. victim or witness. 
In addition, the discovered inconsistency in the responses from the dispatchers also points to the need for designs that will assist the dispatchers in incorporating more standardized question-asking strategies. 
Although not yet implemented by many safety organizations, clear and standardized protocols can be established as tools that can be used by dispatchers during incident handling \cite{clawson2018litigation}. A CA can be designed as a reminder and guide to help dispatchers follow a standardized question-asking protocol, thus reducing dispatchers' training cost \cite{casillo2020chatbot} and providing additional quality assurance.

\textbf{Enhancing CAs' Emotional Awareness} 
During our analysis for RQ2, we found that dispatchers tended to use generic languages with similar or sometimes even repetitive patterns to express emotional support for users. Although these languages served as a gesture of caring and empathy, their generic and repetitive nature might be perceived by users as inauthentic, which can undermine the emotional support provided by the dispatchers \cite{kraak2017authentic}. 
The lack of emotional awareness in CAs might lead to difficulty in establishing rapport with users and awkwardness during conversations \cite{minhas2022protecting, zheng2021pocketbot}. An emotion-aware CA can quickly gain users' trust and elicit more information from them \cite{lee2020hear}, thus significantly improving the efficiency of information gathering during safety incidents. 
Displaying emotional support can also help users to calm down during emergency situations. The CA should be emotionally aware of the user's reactions and attempt to calm the user if necessary, particularly in medical emergencies. The CA should also be able to offer emotional support as well as information about relevant services to assist the caller, but meanwhile be aware of the minimal support that ought to be provided regardless of users' emotional displays and provide reminders to ensure the standard quality of support.

In addition, our findings of RQ2 also suggested an inconsistency in dispatchers' use of tone and responsiveness when handling users' tips. We also found a variety of strategies were adopted by dispatchers in providing emotional support to users. As discussed in Tracy et al.'s work \cite{tracy1998emotion}, 911 emergency call-takers often need to invest emotional labor and experience psychological burnout during the communication with callers. 
For dispatchers, having an emotionally aware CA can also help mitigate the emotional burnout and negative effect of emotional labor by providing suggested candidate messages and automating repetitive tasks, thus improving the efficiency and job satisfaction of dispatchers \cite{harvey2017can}. 
We argue that future studies to incorporate methods from recent related works on empathetic \cite{ma2020survey} and social chatbot \cite{shum2018eliza} to increase the emotional awareness of CA designs.


\textbf{Comprehending Conversational Context Information} 
From our findings of RQ1, a case was observed where the dispatcher was found to have neglected contextual information initially provided by the user when reporting on behalf of the victim. Meanwhile, we discovered that most of the dispatchers' questions contained references that subtly alluded to contextual information expressed in earlier conversations (e.g., "Do you know who the someone was?", "do you know what dorms?"). Context information often plays a crucial role during conversational interactions of incident reporting.
Previous studies have shown that an increased awareness of contextual information during conversation boosts users' trust in the conversation agent \cite{loveys2020effect, rheu2021systematic}, and improves conversation effectiveness \cite{car2020conversational} and users' satisfaction \cite{purington2017alexa}. Our research indicates the value of context comprehension capabilities for the risk reporting conversation agent due to the discovered richness of contextual information during actual reporting conversations. The CA needs to be capable of adapting tone and strategies based on different situations according to conversation contexts.
Recent natural language generation (NLG) research has advanced itself in the direction of increasing the context awareness of task-oriented dialogue systems \cite{wu2020tod, liu2021context, kale2020template}. Future CA design could benefit from improved context awareness by incorporating techniques from related NLP works into the safety reporting domain.

%% file: 6_limitation.tex
\section{Limitations and Future Work}
 Our conversational log analysis uncovers new and significant information but lacks theoretical groundings about users' and dispatchers' reporting behavior. Future research should build on our empirical findings and provide greater insight into the data's surprising patterns.
 During data pre-processing, we only chose the top-6 categories with the highest number of chat utterances. This might limit the generalizability of the downstream analyses. In order to examine how dispatchers collect information from users through conversational interaction, we only investigated conversations with more than 4 utterances.
 We also filtered out tips that are non-English. Although we removed these tips for the simplification of analysis, it is possible that these tips may contain important information that can yield additional insights about risk reporting. Future work should also consider conducting further content analysis over these tips.
 For the annotated data sample, we only investigated two event categories including \textit{TheftLostItem} and \textit{HarassmentAbuse}. This simplifies the annotation process but could leave out important findings that can be derived from the conversation logs under other event categories. Future work needs to be done to further examine the data from the remaining event categories in order to gain a more holistic understanding of text-based reporting system interactions.


%% file: 7_conclusion.tex
\section{Data Collection and Ethics}
\textcolor{blue}{The data was collected and analyzed based on the premise of an NDA (non-disclosure agreement) established between researchers and their institution. The study examines information from organizations and individual users who have consented to the use of anonymized and aggregated data for system improvement research.
The data was carefully anonymized through a number of stages before being accessible to the scholars and used for the study. All personal and sensitive information has been removed, and identifiable information, including user, geolocation, and organization information, has been removed or anonymized prior to researchers gaining access to the data. The organizations and users are represented using anonymous IDs, and all mentions about geolocations, times and person/organization names in the chat log text are detected and replaced with tags (e.g., [LOCATION], [NAME]). The anonymization process renders the dataset hard to be deanonymized.
As an additional measure of anonymization, a random selection of universities was further removed entirely from the dataset by LiveSafe.}

\section{Conclusion}
In this paper, we examined the conversation logs spanning a two-year period from the LiveSafe text-based live chat risk reporting system. Using \textcolor{blue}{quantitative and qualitative} analysis, we investigated how users reported incidents by submitting tips with event descriptions, how administrators from safety organizations responded to users by asking questions about additional event information, and how users' engagement varied based on dispatchers' responses.
We provided empirical evidence strengthening the significance for safety organizations to increase service quality as a method to improve users' risk reporting responsiveness and engagement\textcolor{blue}{, extending the scope from traditional emphasis over effectiveness and efficiency to also valuing emotional support. 
Additionally, we discussed the design features of text-based reporting systems, and the possibilities for creating conversational agents to augment the report handling process, with the goal of providing implications for future text-based risk reporting system design.}

%% file: appendix.tex
\clearpage
\appendix
\onecolumn
\section{APPENDIX}
\label{appendix}

\subsection{Definitions of the Ontology}

\begin{table}[h!]
\resizebox{\linewidth}{!}{
\begin{tabular}{llp{10cm}}
\hline \hline
\multicolumn{3}{c}{\textbf{Definition of Event Entities}}                                                                                                                       \\ \hline \hline
\multicolumn{1}{l|}{\textbf{Entity}}              & \multicolumn{1}{l|}{\textbf{SubEntity}}   & \textbf{Definition}                                                              \\ \hline
\multicolumn{1}{l|}{Person}                       & \multicolumn{1}{l|}{---}                    & A single person or a group of people                                             \\ \hline
\multicolumn{1}{l|}{Location}                     & \multicolumn{1}{l|}{---}                    & A location denoted as a point such as in a postal system or abstract coordinates \\ \hline
\multicolumn{1}{l|}{Weapon}                       & \multicolumn{1}{l|}{---}                    & The primary method/instrument used by the offender which causes harm             \\ \hline
\multicolumn{1}{l|}{Time}                         & \multicolumn{1}{l|}{---}                    & The time an incident occurred                                                    \\ \hline
\multicolumn{1}{l|}{Object}                       & \multicolumn{1}{l|}{---}                    & An inanimate object involved in the incident                                     \\ \hline
\multicolumn{1}{l|}{\multirow{9}{*}{Description}} & \multicolumn{1}{l|}{Person-Age}           & The age of the person                                                            \\ \cline{2-3} 
\multicolumn{1}{l|}{}                             & \multicolumn{1}{l|}{Person-Race}          & The racial description of the person                                             \\ \cline{2-3} 
\multicolumn{1}{l|}{}                             & \multicolumn{1}{l|}{Person-Appearance}    & The physical appearance of the person                                            \\ \cline{2-3} 
\multicolumn{1}{l|}{}                             & \multicolumn{1}{l|}{Person-Clothing}      & The clothing worn by the person                                                  \\ \cline{2-3} 
\multicolumn{1}{l|}{}                             & \multicolumn{1}{l|}{Person-Sex}           & The person's biological sex description                                          \\ \cline{2-3} 
\multicolumn{1}{l|}{}                             & \multicolumn{1}{l|}{Person-Action}        & An action or activity carried out by the person                                  \\ \cline{2-3} 
\multicolumn{1}{l|}{}                             & \multicolumn{1}{l|}{Person-Name}          & The name of the person                                                           \\ \cline{2-3} 
\multicolumn{1}{l|}{}                             & \multicolumn{1}{l|}{Person-Movement}      & The movement or change in location of the person                                 \\ \cline{2-3} 
\multicolumn{1}{l|}{}                             & \multicolumn{1}{l|}{Location-Description} & The descriptive information about the location                                   \\ \hline
\multicolumn{1}{l|}{\multirow{2}{*}{Contact}}     & \multicolumn{1}{l|}{PhoneNumber}          & A phone number                                                                   \\ \cline{2-3} 
\multicolumn{1}{l|}{}                             & \multicolumn{1}{l|}{Email}                & An email address                                                                 \\ \hline
\\ \\
\hline \hline
\multicolumn{3}{c}{\textbf{Definition of Event Arguments}}                                                                                                                      \\ \hline \hline 
\multicolumn{2}{l}{Argument}                                                                  & \textbf{Definition}                                                              \\ \hline
\multicolumn{2}{l|}{ATTACKER}                                                                 & The attacking/instigating agent                                                  \\ \hline
\multicolumn{2}{l|}{TARGET}                                                                   & The target of the offense (including unintended targets)                         \\ \hline
\multicolumn{2}{l|}{WEAPON}                                                                   & The weapon used in the offense                                                   \\ \hline
\multicolumn{2}{l|}{START-TIME}                                                               & When the incident starts                                                         \\ \hline
\multicolumn{2}{l|}{END-TIME}                                                                 & When the incident ends                                                           \\ \hline
\multicolumn{2}{l|}{PLACE}                                                                    & Where the incident takes place                                                   \\ \hline
\multicolumn{2}{l|}{TARGET-OBJECT}                                                            & The target object of the offense (e.g., vehicle, stolen items, etc.)             \\ \hline
\\ \\
\hline \hline
\multicolumn{3}{c}{\textbf{Definition of Dispatchers' Intents}}                                                                                                                                                                                                        \\ \hline \hline
\multicolumn{2}{l|}{\textbf{Intent}}                                                          & \textbf{Definition}                                                                                                                                               \\ \hline
\multicolumn{2}{l|}{Thank}                                                                    & Expressing gratitude to the user                                                                                                                                  \\ \hline
\multicolumn{2}{l|}{ConfirmSendOfficer}                                                       & Confirming an officer has been dispatched                                                                                                                         \\ \hline
\multicolumn{2}{l|}{NotifyOthersInCharge}                                                     & Notifying other responsible personels/parties about the incident                                                                                                  \\ \hline
\multicolumn{2}{l|}{AskMeetOfficer}                                                           & Asking the user to meet an officer offline                                                                                                                        \\ \hline
\multicolumn{2}{l|}{AskToCall}                                                                & Asking the user to call an officer/the organization                                                                                                               \\ \hline
\multicolumn{2}{l|}{AskToVisit}                                                               & Asking the user to visit the organization's office                                                                                                                \\ \hline
\multicolumn{2}{l|}{AskForDetail}                                                             & \begin{tabular}[c]{@{}l@{}}Asking the user for additional detail related to the incident \\ (annotated with the event argument the dispatcher asks about, \\ i.e., ATTACKER, TARGET, WEAPON, START-TIME, \\ END-TIME, PLACE, TARGET-OBJECT)\end{tabular} \\ \hline
\end{tabular}
}
\end{table}